\documentclass[10pt,a4paper,twocolumn]{article}

\usepackage[a4paper,left=1.8cm,right=1.8cm,top=2.2cm,bottom=2.4cm,columnsep=0.7cm]{geometry}

\usepackage{graphicx}%
\usepackage{multirow}%
\usepackage{amsmath,amssymb,amsfonts}%
\usepackage{amsthm}%
\usepackage{mathrsfs}%
\usepackage[title]{appendix}%
\usepackage[dvipsnames,table]{xcolor}%
\usepackage{textcomp}%
\usepackage{manyfoot}%
\usepackage{booktabs}%
\usepackage{algorithm}%
\usepackage{algorithmicx}%
\usepackage{algpseudocode}%
\usepackage{listings}%
\usepackage{tikz}
\usepackage{comment}
\usepackage{subcaption}
\usepackage{caption}
\graphicspath{{./Figures/}}

\definecolor{DarkBlackGreen}{RGB}{30,80,00}

\usepackage{fancyhdr}
\usepackage{titlesec}
\usepackage{cuted}
\usepackage[font=small,labelfont=bf,labelsep=period,justification=justified]{caption}
\usepackage{authblk}
\usepackage{cite}
\usepackage[colorlinks=true,linkcolor=RoyalBlue,citecolor=RoyalBlue,urlcolor=RoyalBlue,breaklinks=true]{hyperref}

\theoremstyle{plain}%
\theoremstyle{definition}%

\titleformat{\section}{\normalfont\large\bfseries}{\thesection.}{0.5em}{}
\titleformat{\subsection}{\normalfont\normalsize\bfseries}{\thesubsection.}{0.5em}{}
\titleformat{\subsubsection}{\normalfont\normalsize\itshape}{\thesubsubsection.}{0.5em}{}
\titlespacing*{\section}{0pt}{1.3em plus 0.3em minus 0.2em}{0.7em}
\titlespacing*{\subsection}{0pt}{1.0em plus 0.2em minus 0.2em}{0.5em}
\titlespacing*{\subsubsection}{0pt}{0.8em plus 0.2em minus 0.2em}{0.4em}

\fancypagestyle{firstpage}{%
  \fancyhf{}%
  \fancyfoot[C]{\small\thepage}%
}

\begin{document}

\twocolumn[
\begin{center}
{\LARGE\bfseries Thermal Control of Hysteresis and Deterministic Chaos in a Memristive MEMS Resonator\par}
\vspace{0.9em}
{\large
N.G.~Koudafok\^e$^{1,2,\ast}$,\;
Thierry~Njougouo$^{3}$,\;
Hilda~A.~Cerdeira$^{1}$,\;
C.H.~Miwadinou$^{4}$
\par}
\vspace{0.6em}
{\small
$^{1}$ICTP South American Institute for Fundamental Research, Instituto de F\'{i}sica Te\'{o}rica, Bloco~II, Rua Dr.\ Bento Teobaldo Ferraz 271, Barra Funda, S\~ao Paulo, 01140-070, Brazil\\
$^{2}$Institut de Math\'ematiques et de Sciences Physiques (IMSP), Universit\'e d'Abomey Calavi (UAC), Dangbo, Porto-Novo, Ou\'em\'e, B\'enin\\
$^{3}$IMT School for Advanced Studies Lucca, Piazza San Francesco 19, 55100 Lucca, Italy\\
$^{4}$\'Ecole Normale Sup\'erieure de Natitingou, Universit\'e Nationale des Sciences, Technologies, Ing\'enierie et Math\'ematiques d'Abomey, B\'enin\\[0.3em]
$^{\ast}$Corresponding author: \texttt{gilles.koudafoke@ictp-saifr.org}
\par}
\end{center}
\vspace{0.8em}
{\rule{\textwidth}{0.6pt}}
\vspace{0.4em}

\noindent{\bfseries Abstract.\ }%
We investigate the nonlinear dynamics of a thermo-electro-mechanically
coupled memristive resonator comprising a doubly clamped Euler--Bernoulli
microbeam, an RLC circuit, and a TiO$_2$ memristor with temperature-dependent
ionic mobility governed by Mott and Efros--Shklovskii hopping conduction. The
dynamics are analyzed using two-dimensional parameter-space maps, bifurcation
diagrams, Lyapunov exponents, reconstructed attractors, Poincar\'e sections,
Grassberger--Procaccia correlation-dimension analysis, empirical mode
decomposition, the Hilbert--Huang spectrum, and electro-memristive hysteresis.
Parameter-space maps reveal predominantly quasi-periodic and deterministic
chaotic regimes without stable phase-locked periodic states. Bifurcation
analyses show that the beam length and excitation frequency govern the dynamics
through the frequency ratio $r_\omega=\omega_0/\omega_b$, whereas the
excitation current mainly controls the oscillation amplitude and chaotic
intensity. Under fixed operating conditions, the asymptotic regime depends on
the initial conditions, and complementary diagnostics identify the
thermo-memristive subsystem as the primary source of the nonlinear complexity,
subsequently transmitted to the microbeam through electromechanical coupling.
Temperature continuously reorganizes the electro-memristive hysteresis through
the chain $T \to \sigma(T) \to M(w,T) \to i_m(t) \to w(t)$. The hysteresis area
evolves non-monotonically with temperature, revealing a
configuration-dependent optimal thermo-memristive operating point. These
findings highlight temperature, beam length, and electrical excitation as
complementary control parameters for tailoring thermo-memristive memory,
deterministic chaos, and nonlinear dynamics in thermo-active MEMS, with
potential applications in neuromorphic sensing and chaos-based secure
communication.

\vspace{0.5em}

\vspace{0.4em}
{\rule{\textwidth}{0.6pt}}
\vspace{1.2em}
]
\thispagestyle{firstpage}


\section{Introduction}
Micro-electromechanical systems (MEMS) have emerged as key platforms for the
development of miniaturized devices integrating sensing, actuation, and signal
processing functions \cite{Ekinci1,Ekinci2,Etaki,Younis2011,Farokhi2015}. Owing to their intrinsic sensitivity to nonlinear and multiphysical effects, MEMS provide a
natural framework for exploring the interplay between mechanical, electrical,
and thermal phenomena \cite{nayfeh1979,shaw1983}. In particular, microbeams described by Euler--Bernoulli theory can exhibit
Duffing-type nonlinearities at large vibration amplitudes, leading to
amplitude-dependent resonance and bistability
\cite{nayfeh1979,holmes1979}. Under appropriate excitation conditions,
MEMS resonators exhibit rich nonlinear dynamical behaviors, including
quasi-periodicity, deterministic chaos, and attractor coexistence, providing
attractive prospects for sensing, nonlinear signal processing, and frequency
tuning \cite{strogatz2015}.

In parallel, the memristor, introduced as the fourth fundamental circuit
element \cite{chua1}, has attracted sustained interest due to its intrinsic
memory and its connection to nonlinear dynamics
\cite{Makoto,wang2015,Xie}. Memristive devices exhibit characteristic pinched
current--voltage hysteresis loops associated with the evolution of internal
state variables \cite{chua1,hadis}, opening perspectives in neuromorphic
electronics, information storage, and adaptive systems
\cite{hadis,Mutlu,mazady,hamdi,Ielmini2016,zidan2018,mokhtar,Belykh,talon}.
While classical memristors are governed by first-order internal dynamics,
higher-order memristive systems incorporating inertia and nonlinear coupling
\cite{diventra2009} display enriched behaviors such as frequency-dependent
hysteresis deformation, self-oscillations, and multi-valued current--voltage
characteristics
\cite{potter2024,pershin2011,muthuswamy2010,buscarino2012,kokate}. The recent
experimental observation of pinched hysteresis in MEMS resonators
\cite{uka2025} demonstrates that mechanical systems can intrinsically exhibit
memory signatures, suggesting that memristive behavior may emerge as a
universal property of coupled dynamical systems. At the same time,
memristive technologies are rapidly evolving toward multifunctional platforms
integrating memory, sensing, and computation
\cite{yarragolla2024,acsnano2023,im2020}. These developments call for
theoretical frameworks that explicitly incorporate inertia, nonlinearity, and
multi-physical coupling.

Despite these advances, the role of temperature in electro-mechanically coupled
memristive systems has not yet been explicitly addressed, although temperature
strongly affects electrical and ionic transport and, in principle, the
mechanical response as well. In resistive switching devices, it governs
ionic mobility, filament formation, and conductivity \cite{waser2007}; in
TiO$_2$-based memristors, conduction proceeds through thermally activated Mott
and Efros--Shklovskii variable-range hopping
\cite{singh2018temperature,khan2024,garcia2016spice}, so that the ON- and
OFF-state resistances and the vacancy mobility inherit a strong temperature
dependence. Here, we address this issue by introducing a coupled
thermo-electro-mechanical model in which the internal state variable of an HP
memristor is controlled by thermally activated ionic transport
\cite{strukov,radwan,kavehei,Makoto}, the resulting nonlinearity being
transmitted to the microbeam through electromechanical coupling. In contrast
to conventional approaches, which primarily consider external forcing of the
mechanical subsystem \cite{Ekinci1,Koudaf,Koudaf2,Koudaf3}, the present
framework reveals that internally controlled thermo-memristive transport
plays a structuring role in the global system dynamics.

The main findings of this work can be summarized as follows. First,
Lyapunov--Benettin maps computed in the $(D,L)$, $(f_0,I_0)$, and $(f_0,L)$
parameter spaces show that the dynamics are predominantly quasi-periodic or
deterministically chaotic, while thermal loading and electrical excitation
continuously reorganize the distribution of nonlinear regimes across the
parameter space. Second, under fixed operating conditions the asymptotic regime is initial-state dependent, and complementary diagnostics---reconstructed attractors, Hilbert--Huang spectra, Poincar\'e sections, and
Grassberger--Procaccia correlation dimensions---identify the thermo-memristive subsystem as the primary source of the nonlinear complexity, transmitted to the microbeam through the electromechanical coupling. Third, the effects of the beam length and the excitation frequency
on the bifurcation structure are shown to be governed by the common frequency
ratio $r_\omega=\omega_0/\omega_b$, which couples the structural geometry to
the electrical forcing, whereas the excitation current mainly controls the
oscillation amplitude and the local chaotic intensity. Finally, temperature
does not trigger chaos but continuously reorganizes the electro-memristive
hysteresis through the thermo-memristive coupling chain
$T \to \sigma(T) \to M(w,T) \to i_m(t)
\to w(t)$,
thereby providing an efficient means of tuning the memristive memory. The
enclosed hysteresis area exhibits a non-monotonic evolution, revealing an
optimal thermo-memristive operating regime near
$T\approx407\,\mathrm{K}$ for a beam length of
$L=30\times10^{-5}\,\mathrm{m}$. This separation of roles between beam length
(structural design), temperature (thermal regulation), and electrical
excitation (dynamic forcing) provides practical control strategies for
engineering memristive and chaotic dynamics in thermo-active MEMS, with
potential applications in neuromorphic sensing, nonlinear signal processing,
and chaos-based secure communication.

The remainder of this paper is organized as follows.
Sec.~\ref{sec2} introduces the thermo-active MEMS structure, its coupled
thermo-electro-mechanical model, and the parameter-space Lyapunov analysis.
Sec.~\ref{sec3} provides evidence for the memristive origin of the
nonlinear complexity and its transfer to the mechanical subsystem, and
reports the initial-state dependence of the asymptotic regime. Sec.~\ref{sec4} examines the excitation- and geometry-induced
bifurcation structures and identifies the frequency ratio
$r_{\omega}=\omega_0/\omega_b$ as the organizing parameter of the nonlinear
dynamics. Sec.~\ref{sec5} addresses the influence of temperature on the
nonlinear dynamics and the thermal modulation of the electro-memristive
hysteresis. Sec.~\ref{sec6} concludes the paper and outlines future research
directions.

\section{Thermo-active MEMS Structure, Mathematical Modeling, and Nonlinear Dynamical Analysis}\label{sec2}
\begin{figure}[!ht!]
\centering
  \includegraphics[width=\columnwidth]{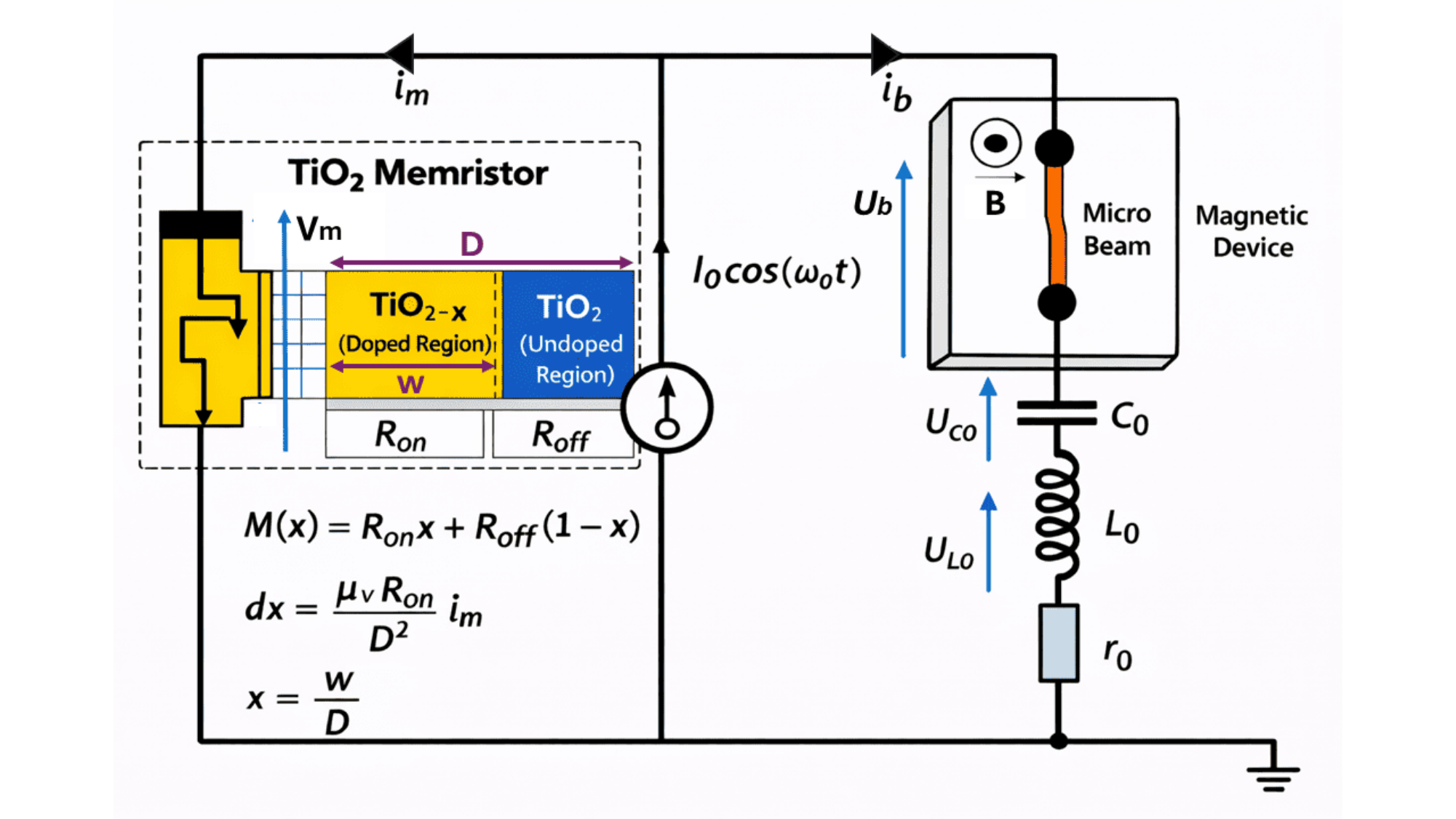}
  \vspace{-0.2cm}
  \caption{\small{Schematic of the memristive MEMS circuit investigated.}} \label{Mems1}
\end{figure}

In this section, we establish the mathematical model used to investigate the MEMS depicted in Fig.~\ref{Mems1}.
The circuit consists of an alternating current source with amplitude $I_0$ and excitation frequency $\omega_0$, a TiO$_2$ memristor, a mechanical resonator (microbeam), and an electrical resonator composed of a resistive inductor $(L_0, r_0)$ and a capacitor $C_0$. The AC source provides energy to the system, while the electrical resonator $(L_0, C_0)$ ensures sustained energy exchange through periodic charging and discharging.
The memristor and the microbeam are used to capture certain physical phenomena such as magnetic fields and temperature, given their hypersensitivity to environmental conditions.
The set of differential equations governing the electrodynamic behavior of the system is derived from Kirchhoff's laws, the electrical transition relations shown in \cite{chua1,chua1976,chua2010,chua2014,strukov}, and the classical Euler--Bernoulli beam theory, which is introduced in the following subsection.

\subsection{Temperature-Dependent Charge Carrier Mobility and Beam
Electrical Resistivity}\label{sec2.1}
The temperature-dependent conductivity of the $\mathrm{TiO_2}$ memristor
is modeled as a superposition of Mott and Efros--Shklovskii (ES) variable
range hopping mechanisms~\cite{Yildiz2008,Efros1975,Mott1968}:
\begin{multline}
\sigma(T) = \sigma_{0,\mathrm{Mott}} T^{-2s}
\exp\!\left[-\!\left(\frac{T_{0,\mathrm{Mott}}}{T}\right)^{\!s}\right]\\
+\, \sigma_{0,\mathrm{ES}} T^{-2s}
\exp\!\left[-\!\left(\frac{T_{0,\mathrm{ES}}}{T}\right)^{\!s}\right]
\label{eq:conductivity}
\end{multline}
where the first term accounts for thermally activated hopping between
localized states (dominant at higher temperatures), and the second
captures Coulomb-gap-mediated transport near the Fermi level (dominant
at lower temperatures)~\cite{Efros1975}. The exponent~$s$ characterizes
the hopping mechanism and~$T_0$ reflects the degree of localization and
the density of states (see Tables~\ref{tab:parameters} and~\ref{tab:2} for the complete parametric description).

The mobility of oxygen vacancies is derived from the conductivity
as~\cite{Bak2003}:
\begin{equation}
\mu_v = \frac{\sigma\, M(\mathrm{TiO_2})}{N_A\, e\, n\, \rho_m(\mathrm{TiO_2})}
\label{eq:mobility}
\end{equation}
where $M(\mathrm{TiO_2})$ and $\rho_m(\mathrm{TiO_2})$ are the molar mass
and mass density of titanium dioxide, $n$ is the effective carrier
concentration, $N_A = 6.022\times10^{23}$~mol$^{-1}$ is Avogadro's
number, and $e = 1.602\times10^{-19}$~C is the elementary charge. Since
$\sigma$ is temperature-dependent through Eq.~\eqref{eq:conductivity},
$\mu_v$ inherits this dependence, making it a key parameter in memristor
switching dynamics~\cite{singh2018temperature}. Consequently, the ON- and
OFF-state resistances, defined within the Singh and Raj
framework~\cite{singh2018temperature} as:
\begin{equation}
R_{\mathrm{on}} = \frac{w}{\sigma A_m} \qquad \text{and} \qquad
R_{\mathrm{off}} = \frac{D - w}{\sigma A_m}
\label{eq:Ron}
\end{equation}
are both temperature-dependent and decrease with increasing
temperature~\cite{singh2018temperature}, where $A_m$ is the
cross-sectional area of the memristor and $w$ is the boundary position
between the doped and undoped regions. The temporal evolution of~$w$,
which governs the memristive memory behavior and hysteresis, is given
by~\cite{singh2018temperature,strukov}:
\begin{equation}
\frac{dw}{dt} = \mu_v \frac{R_{\mathrm{on}}}{D}\, i_m(t)
\label{eq:state}
\end{equation}
The electrical resistivity of the microbeam is expressed in terms of the
electron mobility as:
\begin{equation}
\rho = \frac{1}{e\, N_d\, \mu_e(T)}
\label{eq:rho_beam}
\end{equation}
where $N_d$ is the dopant concentration and $\mu_e(T)$ follows the
empirical Arora model~\cite{arora1982mobility}:
\begin{equation}
\mu_e(T) = \mu_{\min} T_n^{\alpha_s}
+ \frac{\mu_0\, T_n^{\beta}}
{1 + \!\left(\dfrac{N_d}{N_{\mathrm{ref}}\, T_n^{\gamma_s}}\right)^{\!\delta}}
\label{eq:arora_mobility}
\end{equation}
where $T_n = T/T_{\mathrm{ref}}$ is the temperature normalized to $T_{\mathrm{ref}} = 300$~K. The
first term accounts for phonon scattering at elevated temperatures; the
second models mobility degradation due to ionized impurity scattering
at high doping concentrations. The empirical parameters $\mu_{\min}$,
$\mu_0$, $N_{\mathrm{ref}}$, $\alpha_s$, $\beta$, $\gamma_s$, and
$\delta$ are taken from~\cite{arora1982mobility} (see Table~\ref{tab:parameters}).

Regarding thermoelastic damping (TED), the thermal relaxation time is
defined as~\cite{Zener1937,Zener1938,Lifshitz2000}:
\begin{equation}
\tau = \frac{h^2}{\pi^2 D_b}
\end{equation}
where $h$ is the beam thickness and $D_b$ the thermal diffusivity. For
the present device ($h = 1.8\,\mu$m), one finds $\omega_b\tau \ll 1$
over the entire frequency range of interest, placing the system well
within the isothermal elastic regime, far from the thermoelastic
resonance condition. In this limit, TED-induced dissipation scales
approximately linearly with temperature~\cite{Lifshitz2000} but remains
negligibly small compared with other sources of uncertainty. The quality
factor is therefore safely approximated as constant over the investigated
temperature range (200--450~K). Similarly, the Young's modulus~$E$ of
the single-crystal silicon beam is treated as temperature-independent,
a well-justified approximation at the considered doping level
($N_d = 10^{22}$~m$^{-3}$), which remains far below the electronic
degeneracy threshold ($\sim 10^{19}$~cm$^{-3}$) at which electronic
contributions to the elastic constants become
significant~\cite{Hall1967,Keyes1967,Lifshitz2000}.

\subsection{Mathematical and Numerical Modeling}

Applying Euler--Bernoulli beam theory~\cite{Timoshenko1955,Nayfeh2004,Pelesko2002,Senturia2001}, together with Kirchhoff's circuit
laws yields the electromechanical governing
equation of the microbeam, Eq.~(\ref{eq8}), which describes its vibration
amplitude under the combined action of magnetic and electrical excitations~\cite{Koudaf,Koudaf2,Koudaf3}:
\begin{equation}
	\ddot{u}(t) = -\left(\dfrac{\lambda}{\rho A} + \dfrac{16\,B^2\,L}{15 \mu\,\rho} \right) \dot{u}(t) -  \dfrac{E I_y {\xi_n}^4}{\rho A}u(t)  + \dfrac{B L}{\rho A}\dot{q}(t)
     \label{eq8}
\end{equation}
where $E$ denotes the Young's modulus of the beam material,
$I_y$ the second moment of inertia of the beam, $\rho$ the volume density, and $L$ the length of the micro-beam; $\vec{B}$ the magnetic field.
The beam resistance is expressed as $r_b = \frac{\mu L}{A}$, with $A$ being the beam cross-sectional area and $\mu$ the resistivity of the micro-beam material, $u(t)$ the transverse displacement function, and $\lambda$ the damping coefficient. Because the physical deflection $u = u_0\,\theta$ is set by a picometer-scale factor $u_0$ against a beam thickness $h = 1.8\ \mu$m ($u/h \sim 10^{-6}$ throughout), the geometric (Duffing) nonlinearity is negligible~\cite{Younis2011,nayfeh1979} in all regimes, regular or chaotic, justifying the linear Euler--Bernoulli approximation and supporting that the observed nonlinear dynamics originate in
the thermo-memristive coupling rather than in the beam.

The voltage $V_m$ at the terminals of a memristor supplied with a current $i_m$ is written as follows \cite{strukov,Xie,kokate,Mutlu}:
\begin{equation}
\begin{cases}
V_m(t) &= M\!\left[q_m(t)\right]\, i_m(t), \\
&=R_{\mathrm{on}}\left[ \tfrac{R_{\mathrm{off}}}{R_{\mathrm{on}}}  - \mu_v \tfrac{(R_{\mathrm{off}} - R_{\mathrm{on}})}{D^2} q_m(t)\right] i_m(t)\, \\
M\!\left[q_m(t)\right]& =
\dfrac{d\phi_m(t)}{dq_m(t)}
= \dfrac{\dot{\phi}_m(t)}{i_m(t)}.
\end{cases}\label{eq9n}
\end{equation}
where $\phi_m$ is the field flow across the memristor; $q_m$ is the charge across the memristor;
		$D$ is the memristor's length;
	    $\mu_v$ is the memristor ionic mobility;
		$R_{\mathrm{off}}$ is the Off state memristor resistance, and
        $R_{\mathrm{on}}$ is the On state memristor resistance.

Using Kirchhoff's voltage law (KVL) $V_m - U_{\mathrm{beam}} - U_{C_0} - U_{L_0} = 0$, Eq.~(\ref{eq11}) models the variation of the charge $q(t)$ of the capacitor $C_0$.
\begin{equation}
	\ddot{q}(t) = \dfrac{1}{L_0}\dot{\phi}_m(t) - \dfrac{r_p +r_0}{L_0}\dot{q}(t) + \dfrac{16 B L}{15 L_0}\dot{u}(t) - \dfrac{1}{L_0C_0}q(t)
    \label{eq11}
\end{equation}
where $r_{0}$ is the Coil resistance;
	   $C_0$ is the Self-capacitance of the capacitor,
	   $L_0$ is the Self-inductance of the coil,
	   $I_{0}$ is the amplitude of the excitation current. \\

From the $\dot{\phi}_m$ equation obtained earlier (Eq.~(\ref{eq9n})), using Kirchhoff's law and differentiating it once as a function of time, we obtain the equation modeling the variation in memristor flux $\phi_m$ at the TiO$_2$ memristor model terminals (see Sec.~\ref{Sec7}, (\ref{eq12})).

While posing~$\tau = \omega_1 t; \, \theta = \dfrac{u(t)}{u_0}; \, \gamma = \dfrac{q(t)}{Q_0}$ \,\, and \,\, $\alpha_m = \dfrac{\phi_m(t)}{\phi_0}$ where $\omega_1, u_0, Q_0, \phi_0$ are scaling coefficients, and $x=\dot{\alpha}_m(\tau); \quad y=\dot{\theta}(\tau); \quad z=\dot{\gamma}(\tau)$, Eqs.~(\ref{eq:state}), (\ref{eq8}), (\ref{eq9n}), (\ref{eq11}) representing the flow governing the dynamic system are as follows:\\
{\footnotesize
\begin{equation}\label{eq15}
	\begin{cases}
			\dot{\alpha}_m(\tau)=x \\
			\dot{x} = - \Delta_{R_m}  \left(I_0\cos(\omega^*_0 \tau) - \omega_1\,Q_0\,z + \sigma_0 \,y\right)^2 \\
			\quad + \sigma_1 \left\lbrace  R^*_0 - \Delta_{R_{\mathrm{on}}} \left[ I^*_0 \sin(\omega^*_0 \tau) - Q_0\,\gamma(\tau) +\sigma_2\theta(\tau)\right] \right\rbrace \\
			\quad \times \left\lbrace -I^*_0\sin(\omega^*_0 \tau)
			- \vartheta_2\,x
			+ \delta_{r_2}\,z
			\right.\\ \qquad \left. -\sigma_3\,y
			+ \sigma_4\gamma(\tau)-
			 \sigma_5\theta(\tau)
			\right\rbrace\\
			\dot{\theta}(\tau)=y\\
			\dot{y}= -\beta_1 y - \beta_2\theta(\tau)  + \beta_3\,z\\
			\dot{\gamma}(\tau)=z\\
			\dot{z} = \vartheta_1\,x - \delta_{r_1}\,z + \eta\,y - \Omega^2_{C_0}\,\gamma(\tau)\\
            \dot{w}(\tau) = \mu_v \dfrac{R_{\mathrm{on}}}{\omega_1 D} i_m(\tau) \\
             i_m(\tau) = \dfrac{\vartheta_0\,x}{  R^*_0 - \Delta_{R_{\mathrm{on}}} \left\lbrace I^*_0 \sin(\omega^*_0 \tau) - Q_0\,\gamma(\tau) +\sigma_2\theta(\tau) \right\rbrace }
	\end{cases}
    \end{equation}
}
with: $\beta_1$, $\beta_2$, $\beta_3$, $\omega_{C_0}^2$, $\Omega_{C_0}^2$, $\omega_0^{*}$, $I_0^{*}$, $R_0^{*}$, $\vartheta_0$, $\vartheta_1$, $\vartheta_2$, $\delta_{r_1}$, $\delta_{r_2}$, $\Delta_{R_{\mathrm{on}}}$, $\Delta_{R_m}$, $\sigma_0$, $\sigma_1$, $\sigma_2$, $\sigma_3$, $\sigma_4$, $\sigma_5$ defined in Sec.~\ref{Sec7}, Table~\ref{System_Param}.\\

\begin{table*}[!t]
\centering
\caption{Physical and geometrical parameters used in the simulations (See \cite{strukov,Ji2021, Mohanty, Etaki, Ekinci1, Ekinci2, singh2018temperature,arora1982mobility}.)}
\label{tab:parameters}
\begin{tabular}{@{}>{\raggedright\arraybackslash}p{0.11\textwidth} p{0.35\textwidth} >{\raggedright\arraybackslash}p{0.11\textwidth} p{0.35\textwidth}@{}}
\toprule
\textbf{Parameter} & \textbf{Description and Value} & \textbf{Parameter} & \textbf{Description and Value} \\
\midrule
$Q$ & Quality factor: $1.0 \times 10^{4}$ & $B$ & Magnetic field: $0.5$~T \\
$L$ & Beam length: $[15-30]\times 10^{-5}$\, m & $l$ & Beam width: $1.2 \times 10^{-6}$ m \\
$h$ & Beam thickness: $1.8 \times 10^{-6}$ m & $A$ & Cross-sectional area: $2.16 \times 10^{-12}$ m$^2$ \\
$N_d$ & Dopant concentration: $1.0 \times 10^{22}$ m$^{-3}$ & $D$ & Memristor length: $[15-30] $ nm \\
$A_m$ & Memristor cross-section: $1.0 \times 10^{-12}$ m$^2$ & $E$ & Young's modulus: $1.5 \times 10^{11}$ Pa \\
$\rho_m$ & Mass density: $2.33 \times 10^{3}$ kg/m$^3$ & $L_0$ & Inductance: $5.0 \times 10^{-2}$ H \\
$C_0$ & Capacitance: $500$\,pF & $r_0$ & Resistance: $5.0~\Omega$ \\
$f_0$ & Excitation frequency: $[15-80]$\,kHz & $\mu_{\min}$
& Minimum mobility at high doping: $\mathrm{88\,cm^2\,V^{-1}\,s^{-1}}$\\
$\alpha_s$ & Empirical fitting parameters: $-0.57$ & $\mu_{0}$
& Low-doping mobility contribution: $\mathrm{1251.81\,cm^2\,V^{-1}\,s^{-1}}$\\
$\beta$ & Empirical fitting parameters: $-2.33$ & $N_{A}$
& Avogadro's number: $6.022 \times 10^{23}\ \mathrm{mol^{-1}}$\\
$\gamma_s$ & Empirical fitting parameters: $2.4$ & $N_{\mathrm{ref}}$
& Reference concentration: $1.26 \times 10^{17}\,\mathrm{cm^{-3}}$\\
\bottomrule
\end{tabular}
\end{table*}

\begin{table*}[!t]
\caption{Mott and ES constants for TiO$_2$ sample (L-31) (See \cite{singh2018temperature} Tab. 1)}\label{tab:2}%
\centering
\begin{tabular}{@{}lllllll@{}}
\toprule
\textbf{Sample} & $\boldsymbol{\sigma_{0,\mathrm{Mott}}}$ ($\Omega\cdot\mathrm{cm})^{-1}$ & $\boldsymbol{T_{0,\mathrm{Mott}}}$ (K) & $\boldsymbol{s_{\mathrm{Mott}}}$ & $\boldsymbol{\sigma_{0,\mathrm{ES}}}$ ($\Omega\cdot\mathrm{cm})^{-1}$ & $\boldsymbol{T_{0,\mathrm{ES}}}$ (K) & $\boldsymbol{s_{\mathrm{ES}}}$ \\
\midrule
L-31 & $1.53 \times 10^{6}$ & $2.41 \times 10^{6}$ & 0.26 & 30.32 & $2.17 \times 10^{2}$ & 0.48 \\
\bottomrule
\end{tabular}
\end{table*}

\subsection{Thermo-Electro-Mechanical Coupling and Nonlinear Dynamics of the System}\label{sec2.2}
The nonlinear governing equations (Eq.~\ref{eq15}) derived in the previous section are now solved numerically to investigate the coupled thermo-electro-mechanical response of the proposed memristive MEMS. The numerical analysis aims to highlight the relationship between thermal loading, excitation parameters, geometrical parameters, and the resulting nonlinear dynamics, thereby providing a comprehensive understanding of the operating regimes of the system.

All numerical analyses are performed in Python using a fourth-order
Runge--Kutta (RK4) scheme with a fixed integration step
$dt = 5\times10^{-4}$ (dimensionless time units). Unless otherwise
specified, trajectories are integrated up to $\tau_f = 5000$; for
Fig.~\ref{fig7}, the integration horizon is extended to
$\tau_f = 10\,000$ to confirm the persistence of the chaotic regime
over an extended time span. In all cases, the first $80\%$ of each
integrated trajectory is discarded as transient before any diagnostic
bifurcation diagram, Lyapunov exponent, reconstructed attractor,
or hysteresis loop is computed on the remaining steady-state
segment. The internal state variable \(w(t)\), representing the width of the oxygen-vacancy-rich region, is physically bounded by the device thickness such that $0 \leq \frac{w(t)}{D} \leq 1$. For the two-dimensional Lyapunov--Benettin maps
(Figs.~\ref{fig3} and~\ref{fig6}), the largest Lyapunov exponent is
estimated using the classical two-trajectory Benettin
algorithm~\cite{Benettin1980a,Benettin1980b}, with an initial perturbation
$\delta_0 = 10^{-8}$ renormalized every $100$ integration steps. Since the system is periodically forced through $\cos(\omega_0^{*}\tau)$,
the broken time-translation invariance implies that the Benettin
two-trajectory estimate returns the largest Lyapunov exponent
$\lambda_{\max}$ of the stroboscopic map: $\lambda_{\max}<0$, $\simeq 0$,
and $>0$ correspond to phase-locked periodic, quasi-periodic, and
chaotic motion, respectively. Regimes are classified with a tolerance
$\varepsilon = 2\times10^{-4}$ fixed by the integration horizon $\tau_f$.

\begin{table*}[!t]
\caption{Table summarizing the nature and unit of the physical parameters of the system used in the numerical study.}\label{tab3}%
\centering
\begin{tabular}{@{}llll@{}}
\toprule
\textbf{Designation} & \textbf{Nature} \& \textbf{Unit} & \textbf{Designation} & \textbf{Nature} \& \textbf{Unit}\\
\midrule
\textbf{$i_{m}$} & Dimensionless memristor current & \textbf{$(\alpha_m;\,\, \dot{\alpha}_m)$} & Dimensionless Memristor variables\\
\textbf{$(\gamma;\,\, \dot{\gamma})$} & Dimensionless $C_0$ charge variables & \textbf{$(\theta;\,\, \dot{\theta})$} & Dimensionless Beam variables\\
 \textbf{$i_{C_0} = \dot{\gamma} \sim i_b$}  & Dimensionless Capacitor $C_0$ current & \textbf{$\tau$} & Dimensionless time\\
\bottomrule
\end{tabular}
\end{table*}

The numerical values of the main system parameters are listed in Tables~\ref{tab:parameters} \& \ref{tab:2}, and the physical meanings of the various variables used in the numerical analyses are listed in Table~\ref{tab3}. The axes carrying the dimensionless variables will remain unitless during the numerical analysis. The initial conditions are chosen as $[x^*,\alpha^*_m,y^*,\theta^*,z^*,\gamma^*,w^*,i_m^*] = [0, 0, 0, 0, 10^{-6}, 0, 0, -5.5]$.

\subsubsection{Coupled thermal and physical parameter effects on the nonlinear internal dynamics and electromechanical response of the system}\label{sec2.2.1}
\begin{figure}[!t]
    \centering
    \begin{subfigure}[b]{0.48\columnwidth}
        \centering
        \includegraphics[width=\textwidth,height=3.0cm]{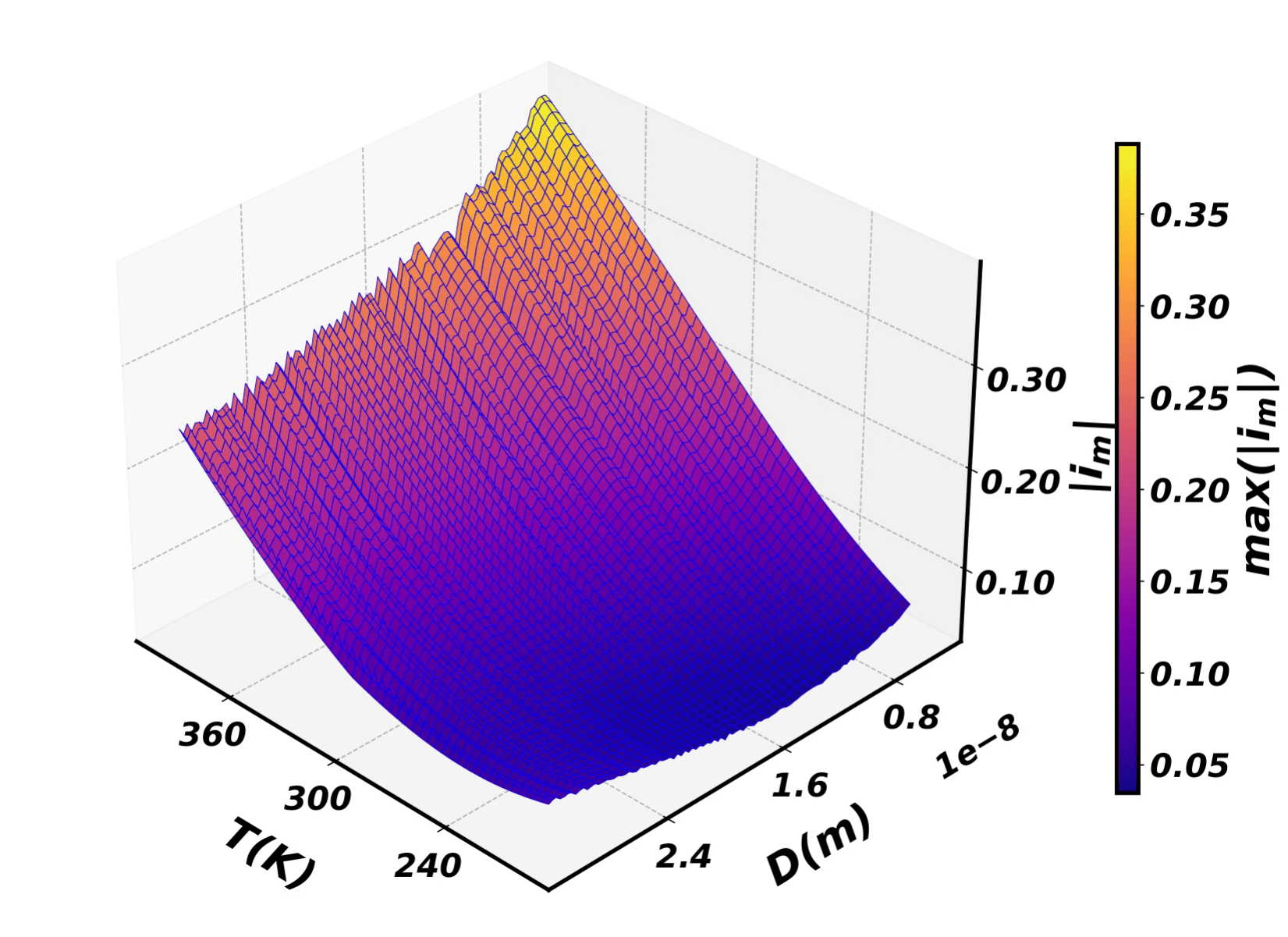}
        \captionsetup{justification=raggedright,singlelinecheck=false}
        \caption{\tiny{Variation of\,\,$i_m = f(T, D)$}}\label{fig2a}
    \end{subfigure}
    \hfill
    \begin{subfigure}[b]{0.48\columnwidth}
        \centering
        \includegraphics[width=\textwidth,height=3.0cm]{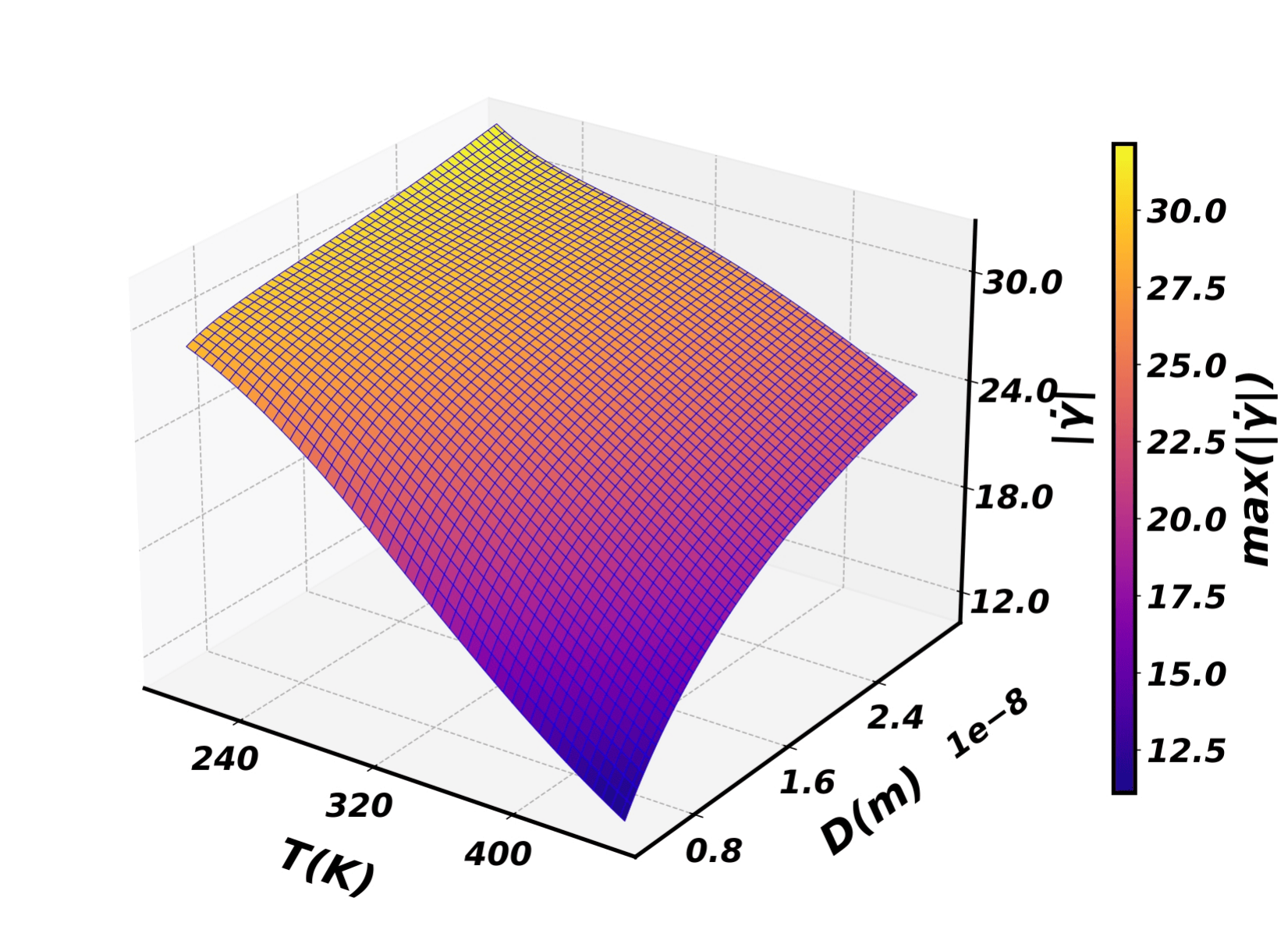}
        \caption{\tiny{Variation of\,\,$i_{b}=f(T, D)$}}\label{fig2b}
    \end{subfigure}

    \vspace{0.3em}
     \begin{subfigure}[b]{0.48\columnwidth}
        \centering
        \includegraphics[width=\textwidth,height=3.0cm]{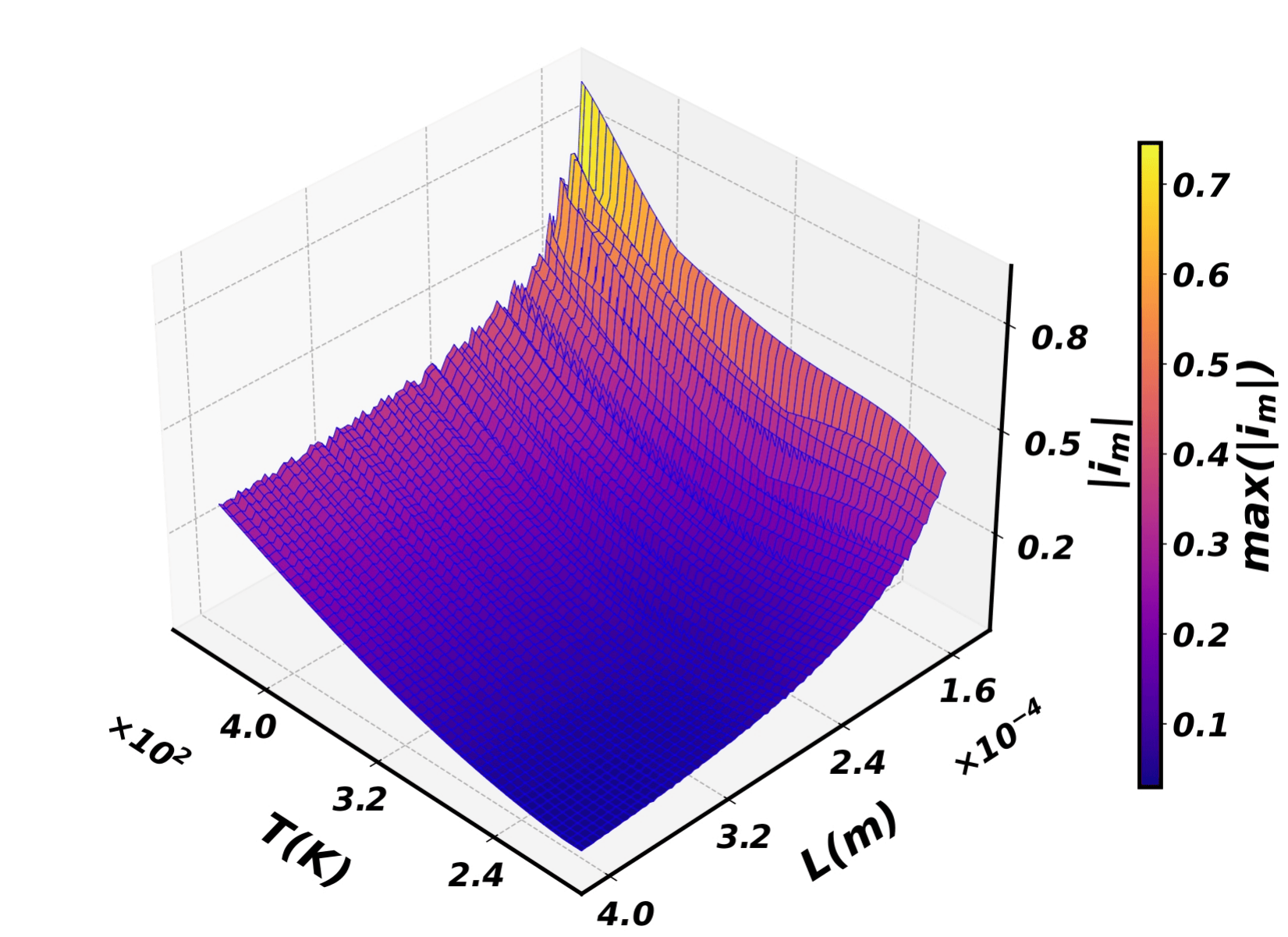}
        \captionsetup{justification=raggedright,singlelinecheck=false}
        \caption{\tiny{Variation of\,\,$i_m = f(T, L)$}}\label{fig2c}
    \end{subfigure}
    \hfill
    \begin{subfigure}[b]{0.48\columnwidth}
        \centering
        \includegraphics[width=\textwidth,height=3.0cm]{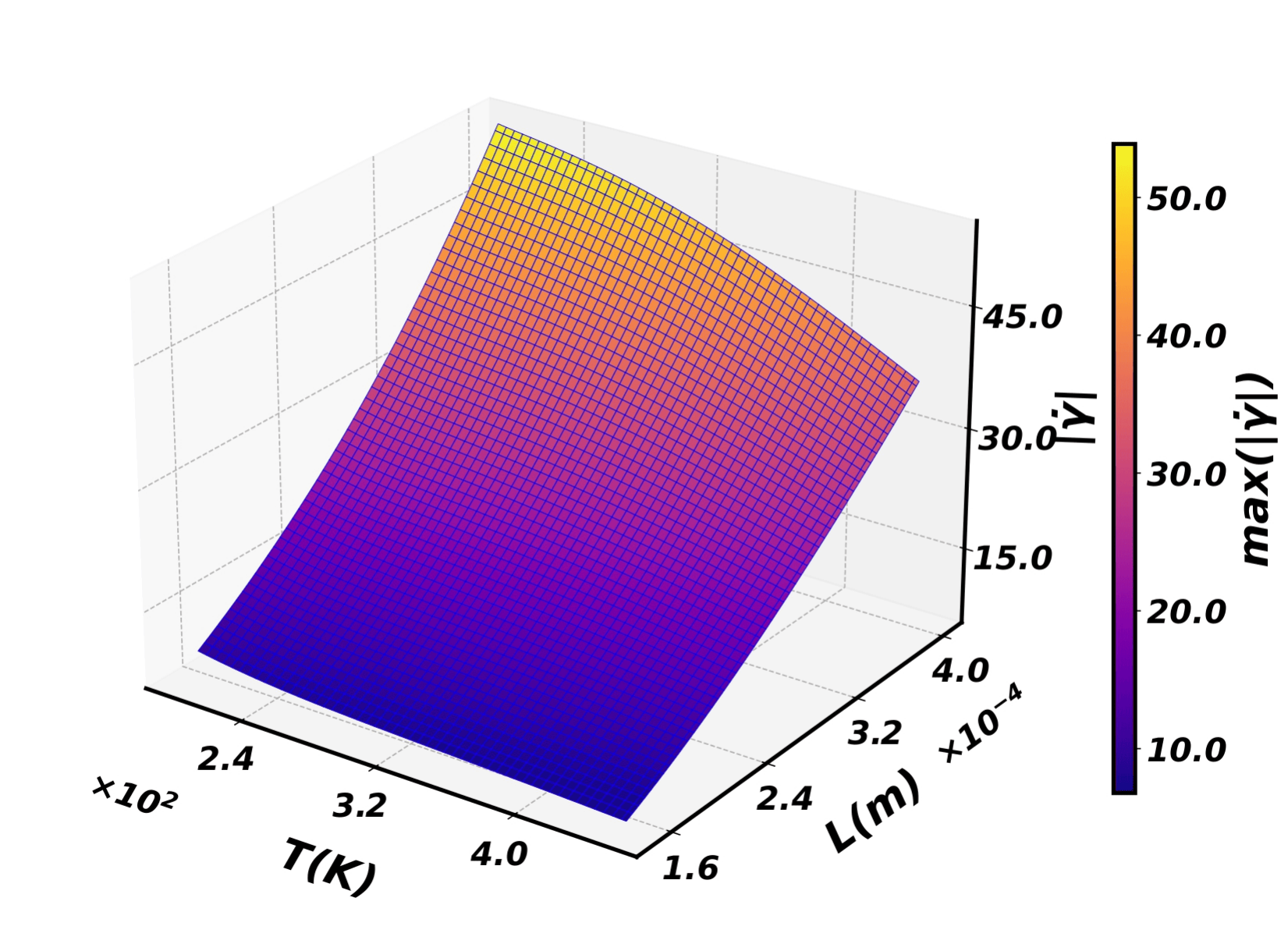}
        \caption{\tiny{Variation of\,\,$i_{b}=f(T, L)$}}\label{fig2d}
    \end{subfigure}
    \caption{\small{3D surfaces of the memristive current $i_m$ and resonator current $i_b$ under thermo-geometrical coupling: (a) The current ($i_m$) through the memristor and (b) The current through ($i_b$) the beam and capacitor $C_0$ as a function of $(T,D)$. (Fig.~\ref{fig2}(b) is rotated with respect to Fig.~\ref{fig2}(a), to appreciate the response to the current). (c) The current ($i_m$) through the memristor and (d) The current through ($i_b$) the beam and capacitor $C_0$ as a function of $T$ and $L$. (Fig.~\ref{fig2}(c) is rotated with respect to Fig.~\ref{fig2}(d), to appreciate the response to the current). $I_0=1\,\mathrm{mA}$ and $f_0 =4.23\,\mathrm{kHz}$.}}\label{fig2}
\end{figure}

Fig.~\ref{fig2} illustrates the evolution of the maximum amplitudes of the memristive currents, $|i_m|$ and $\dot{\gamma}\sim i_b$, as functions of temperature $T$ and the parameters $D$ (memristor length) and $L$ (microbeam length). The results corroborate the consistency of the thermo-electro-mechanical model with Kirchhoff's laws, as evidenced by the inverse relationship between $i_m$ and $i_b$. They further indicate the strong sensitivity of the current amplitudes to thermal loading and parameter variations, highlighting their combined influence on the electromechanical response of the vibrating beam.

\textit{\small{Note: The sub-figures Figs.~\ref{fig2a}, \ref{fig2b}, \ref{fig2c}, \ref{fig2d} are displayed with optimized viewing perspectives to enhance the visualization of the system dynamics; consequently, the axis orientations may differ between panels.}}

The joint analysis of the $|i_m|$ and $|\dot{\gamma}|$ surfaces reveals opposite thermal responses. While $|i_m|$ increases with temperature, $|\dot{\gamma}|$ decreases, reflecting the competition between thermally enhanced memristive conduction and mechanical dissipation. The oscillatory structures observed only on the $|i_m|$ surface further indicate that the memristor is the primary source of the system's nonlinear complexity, exhibiting memory effects and dynamical transitions (see Figs.~\ref{fig2}(a,c) and~\ref{fig3}). Through electromechanical coupling, these nonlinearities are transmitted to the vibrating beam. The increase of $|i_m|$ with temperature is consistent with the thermal enhancement of the effective conductivity $\sigma_m$, which promotes ionic mobility, in agreement with the results of Singh and Raj~\cite{singh2018temperature}.
Overall, these results show that the coupled effects of temperature and the $(D,L)$ parameters provide effective control over the nonlinear electromechanical response of the thermo-active memristive MEMS.

\subsubsection{Parameter-space analyses}\label{sec2.2.2}
\begin{figure}[!t]
\centering
\includegraphics[width=\columnwidth,height=5.0cm]{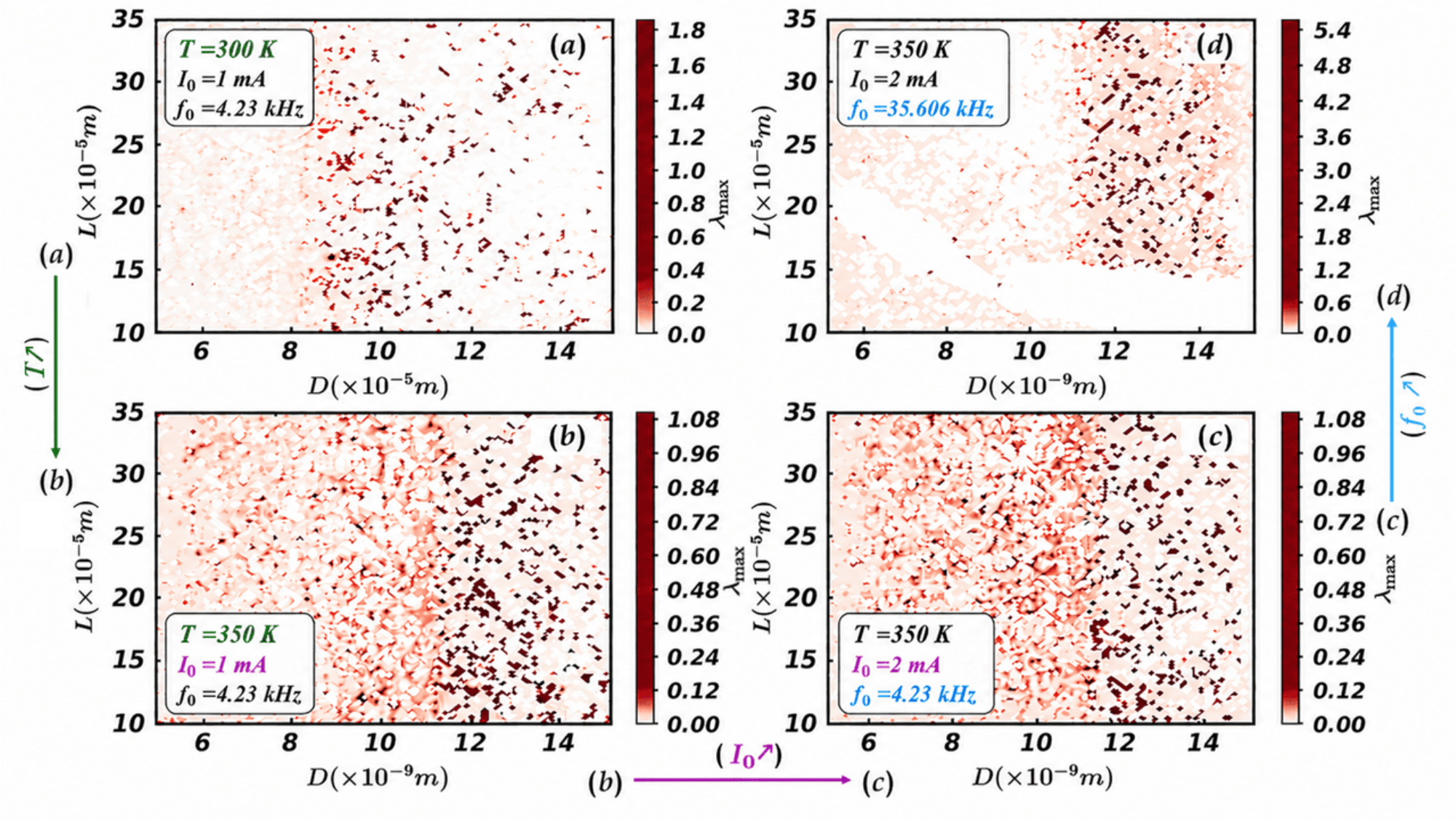}
\vspace{-0.2cm}
\caption{\small{Lyapunov--Benettin maps revealing the distribution of nonlinear dynamical regimes in the $(D,L)$ parameter space.}} \label{fig3}
\end{figure}

To further investigate the influence of the governing physical parameters on the nonlinear dynamics of the proposed thermo-active MEMS system, four maximum Lyapunov exponent diagrams computed using the classical Benettin renormalization algorithm~\cite{Benettin1980a,Benettin1980b} are presented in Fig.~\ref{fig3}.

Fig.~\ref{fig3} presents four Lyapunov--Benettin maps in the $(D,L)$
parameter space, computed from the memristor current $i_m$, illustrating
the combined influence of geometry, excitation conditions, and thermal loading on the nonlinear dynamics. The largest Lyapunov exponent $\lambda_{\max}$ serves as a stability indicator, with $\lambda_{\max}>0$ denoting chaos and $\lambda_{\max}\simeq0$ quasi-periodicity. Throughout the explored parameter space, $\lambda_{\max}$ remains positive or close to zero, indicating the absence of phase-locked periodic regimes. This behavior arises from the coexistence of the excitation ($\omega_0$), mechanical
($\omega_b \simeq (h/L^2)\sqrt{E/\rho}$), and electrical
($\omega_{C_0}=1/\sqrt{L_0C_0}$) frequencies. A periodic response would require the simultaneous commensurability of two independent frequency ratios, a codimension-two condition of negligible measure. Consequently, the dynamics are dominated by quasi-periodic tori that evolve into deterministic chaos, as confirmed by the closed-curve Poincar\'e section for Initial
Condition~1 (Fig.~\ref{fig11}).

The maps reveal a progressive reorganization of the dynamical regimes.
In the reference configuration (Fig.~\ref{fig3}(a)), chaos remains
localized and mainly appears for $D \gtrsim 9$~nm. Increasing the
temperature to $350$~K (Fig.~\ref{fig3}(b)) enlarges and fragments the
chaotic regions, shifting the transition boundary to $D \approx 11$~nm
and confirming the destabilizing role of thermal effects. Raising the
excitation current to $2$~mA (Fig.~\ref{fig3}(c)) further expands and
connects the chaotic domains, reflecting enhanced thermo-electrical
feedback. Finally, increasing the excitation frequency to
$f_0 = 35.606$~kHz (Fig.~\ref{fig3}(d)) produces the most pronounced
reorganization, with a broad diagonal stable band separating highly
chaotic regions. Altogether, these results demonstrate that the
interplay between geometrical parameters, thermal loading, and
electrical excitation governs the global organization of stable and
chaotic regimes in the thermo-active memristive MEMS.

\section{Memristive Origin of Nonlinear Complexity and Initial-State Dependence}
\label{sec3}
\begin{figure*}[!t]
\centering
\includegraphics[width=0.95\textwidth]{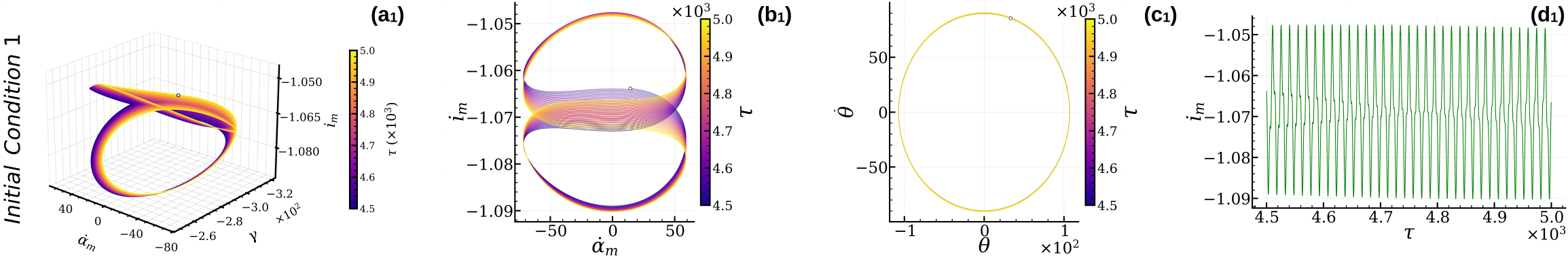}\\[0.2cm]
\includegraphics[width=0.95\textwidth]{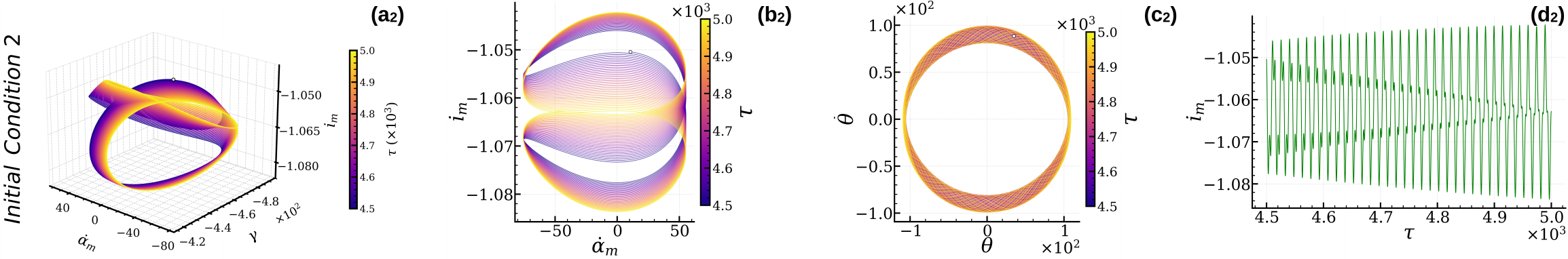}\\[0.2cm]
\includegraphics[width=0.95\textwidth]{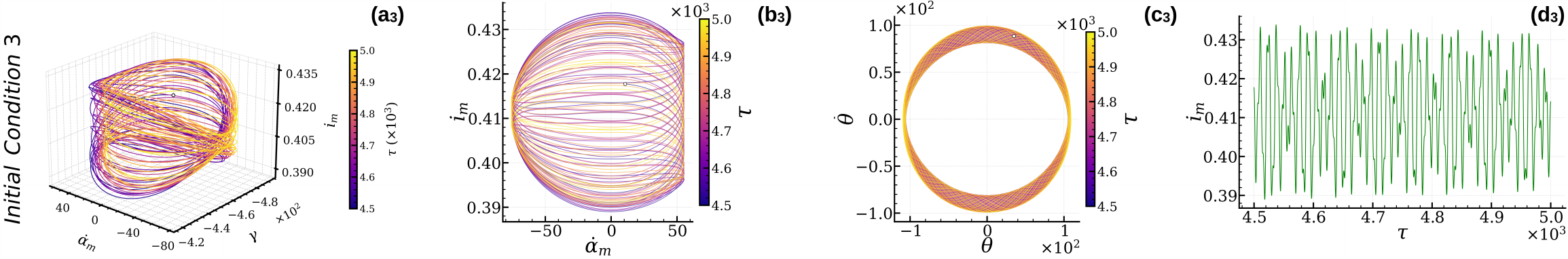}
  \vspace{0.2cm}
  \caption{\small{Initial-state-dependent regime selection under fixed operating
conditions: reconstructed attractors ($a_1$--$a_3$), electro-memristive
phase-space projections ($b_1$--$b_3$), mechanical phase portraits
($c_1$--$c_3$), and memristive-current time responses ($d_1$--$d_3$). Initial
condition~1: $(0,0,0,0,10^{-6},0,0,-5.5)$; initial condition~2: $(0.5,-1,0.25,0.33,10^{-6},0,0,-5.5)$; initial condition~3:
$(0.5,-1,0.25,0.33,10^{-6},0,0,-3.5)$.
Parameters: $T=350$~K, $L=25\times10^{-5}$~m, $D=25$~nm, $I_0=1.2$~mA,
$f_0=38.85$~kHz.}}\label{fig10}
\end{figure*}

The two-dimensional Lyapunov--Benettin maps provide a global description of
the dynamical organization but do not identify \emph{where} the nonlinear
complexity originates, nor whether the asymptotic response is unique for a
given operating condition. To address both questions, all physical
parameters, including the beam length, the excitation frequency, and the
remaining system parameters, are kept strictly fixed, and only the initial
conditions are varied (Fig.~\ref{fig10}). This isolates the
intrinsic response of the coupled system from any parametric effect.

\begin{figure*}[!t]
\centering
  \includegraphics[width=0.95\textwidth]{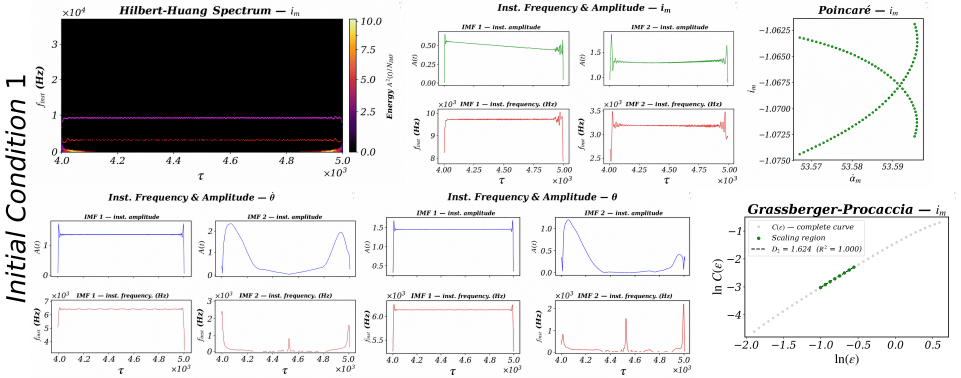}
\vspace{0.35cm}\\
  \includegraphics[width=0.95\textwidth]{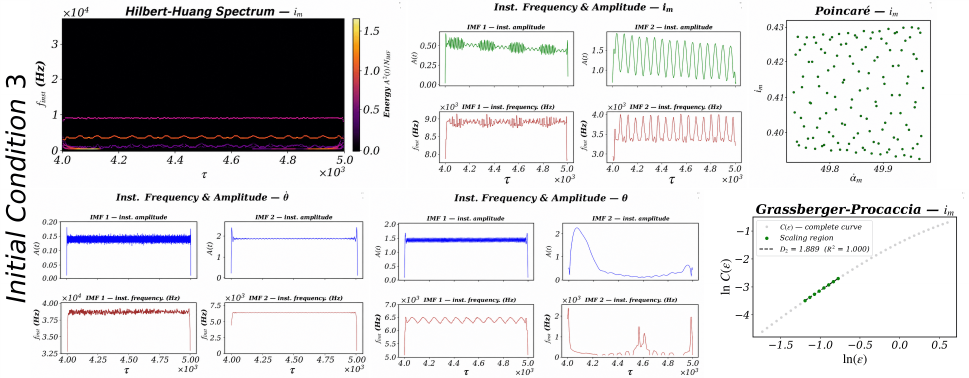}
 \caption{\small{Memristive origin of the nonlinear complexity and its transfer to the mechanical subsystem, characterized by Hilbert--Huang analysis, Poincar\'e sections, and Grassberger--Procaccia correlation dimensions for the regular (initial condition~1) and chaotic (initial condition~3) responses. Parameters:
$T=350$~K, $L=25\times10^{-5}$~m, $D=25\times10^{-9}$~m, $I_0=1.2$~mA,
$f_0=38.85$~kHz.}} \label{fig11}
\end{figure*}

Two observations follow. First, the asymptotic regime is
\emph{initial-state dependent}: identical operating conditions lead either to
a regular, quasi-periodic response or to a low-dimensional deterministic
chaotic one, depending solely on the initial state
(Fig.~\ref{fig10}). We report this dependence as an observed
property of the coupled system; a systematic characterization of the
associated basins of attraction is deferred to future work. Second, and more
importantly, the complementary diagnostics of Fig.~\ref{fig11} reveal the
\emph{direction} of the underlying coupling. For Initial Condition~1, all diagnostics consistently indicate regular dynamics.

The Hilbert--Huang spectrum exhibits a few narrow stationary bands,
the IMFs show slowly varying amplitudes and nearly constant instantaneous
frequencies, and the Poincar\'e section forms a closed curve. The
Grassberger--Procaccia analysis gives a well-defined scaling region
($R^2=1.000$) with $D_2=1.624$ (between a limit cycle, $D_2=1$, and a fully
developed two-torus, $D_2=2$), indicating a weakly modulated quasi-periodic
orbit. These results support a regular, non-chaotic attractor. For Initial
Condition~3, the memristor current $i_m$ becomes irregular first---its
Hilbert--Huang energy spreads over broad, undulating bands, and its IMFs
exhibit strong amplitude modulation and erratically fluctuating instantaneous
frequencies---and this irregularity is subsequently transmitted to the
mechanical variables $\theta$ and $\dot{\theta}$, whose modes lose the smooth
character observed for Initial Condition~1. The Poincar\'e section of $i_m$
becomes a scattered cloud and the correlation dimension rises to
$D_2 = 1.889$ ($R^2 = 1.000$), consistent with a low-dimensional
deterministic chaotic attractor.

This temporal ordering---irregularity emerging in the memristive current
and only then appearing in the mechanical response---identifies the
thermo-memristive subsystem as the primary source of the nonlinear
complexity, the vibrating beam acting as a driven subsystem through the
electromechanical coupling. This interpretation is consistent with the
current-amplitude analysis of Fig.~\ref{fig2}, where the
oscillatory structures appear on the $|i_m|$ surface alone, and it
provides evidence for the memristive origin of the chaos as the central mechanism organizing the coupled dynamics.

\section{Excitation-Induced Bifurcation Structures and Beam-Length Reorganization of Nonlinear Dynamics}\label{sec4}
\begin{figure*}[!t]
    \centering
    \begin{subfigure}[b]{0.33\textwidth}
        \centering
        \includegraphics[width=\textwidth]{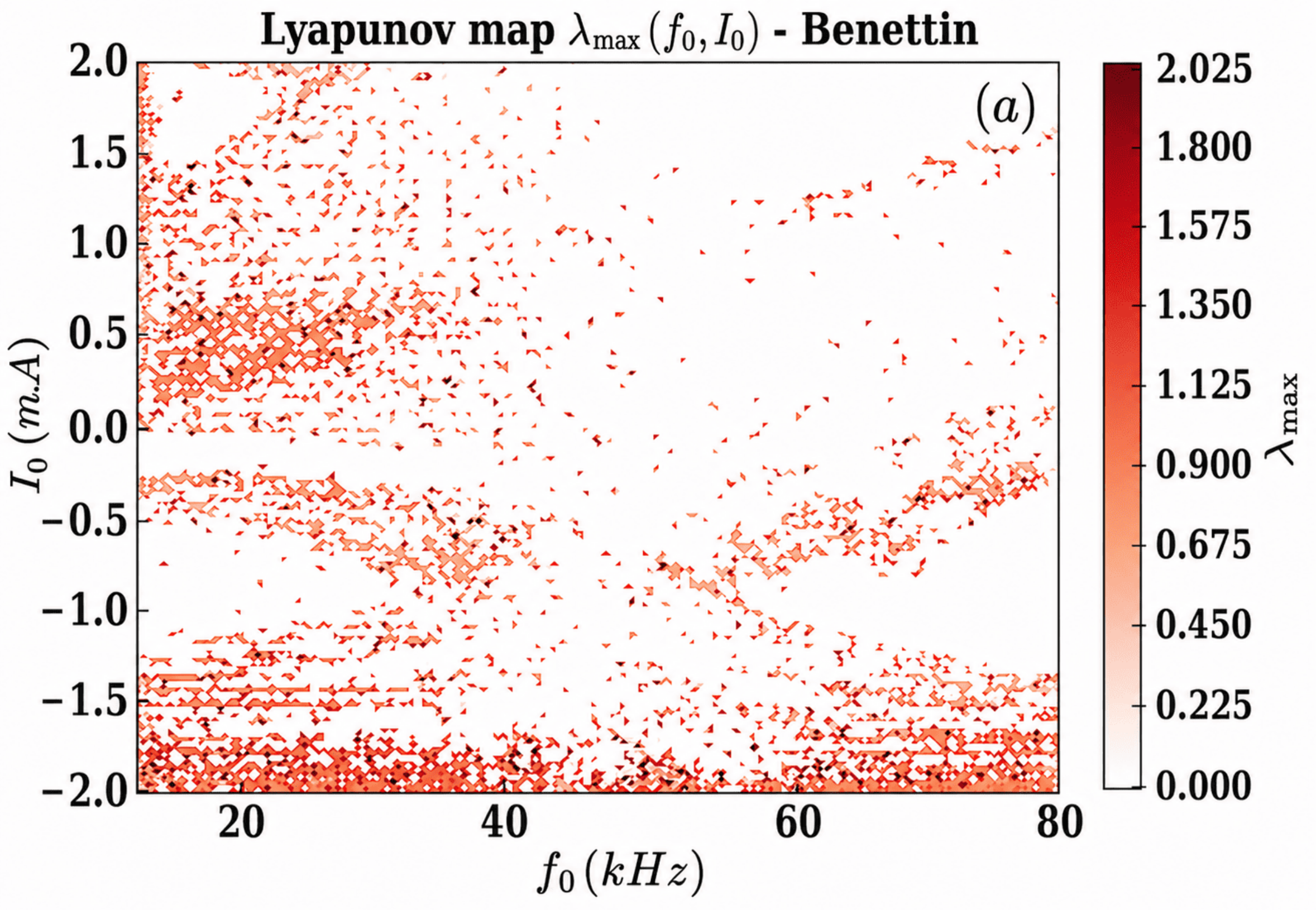}
        \caption{Lyapunov map in the $(f_0,I_0)$ parameter space. \space Parameters: $T=350\,\mathrm{K}$, $L=25\times10^{-5}\,\mathrm{m}$, $D=25\,\mathrm{nm}$.}
        \label{fig4a}
    \end{subfigure}
    \hspace{0.04\textwidth}
    \begin{subfigure}[b]{0.33\textwidth}
        \centering
        \includegraphics[width=\textwidth]{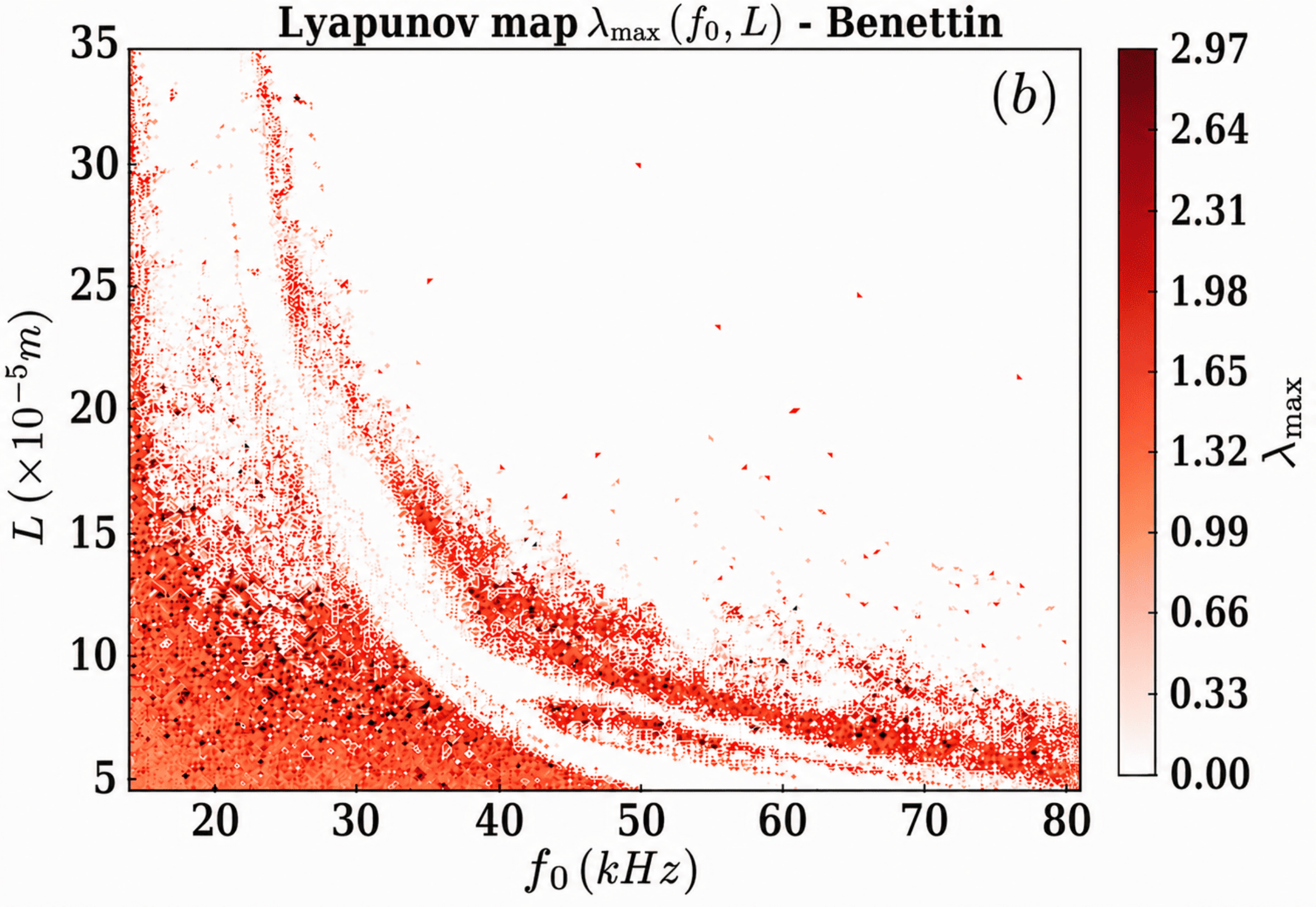}
        \caption{Lyapunov map in the $(f_0,L)$ parameter space. \space Parameters: $T=350\,\mathrm{K}$, $D=25\,\mathrm{nm}$, $I_0 = 1.2\,\mathrm{mA}$.}
        \label{fig4b}
    \end{subfigure}
    \caption{\small{Global dynamical landscapes derived from the maximum Lyapunov exponent of the memristor current $i_m$, illustrating the occurrence of quasi-periodic and chaotic regimes in the $(f_0,I_0)$ parameter space (a) and the $(f_0,L)$ parameter space (b), with no purely periodic regime found throughout the investigated parameter domains.}}
\label{fig4}
\end{figure*}

A global overview of the nonlinear dynamics under the combined influence of the excitation frequency $f_0$, input current $I_0$, and beam length $L$ is provided through two-dimensional dynamical maps constructed in the $(f_0,I_0)$ and $(f_0,L)$ parameter spaces (Fig.~\ref{fig4}).

Unlike one-dimensional bifurcation diagrams, these maps reveal the global organization of the parameter space, highlighting the interplay between excitation frequency, electrical forcing, and structural geometry. They identify regions associated with regular, quasi-periodic, and chaotic dynamics and provide a basis for selecting representative parameter sets for the detailed bifurcation analyses presented in the following sections.
A comparison of Figs.~\ref{fig4}(a) and~\ref{fig4}(b) shows that the beam length $L$ has a stronger influence on the global dynamics than the input current $I_0$. In the $(f_0,L)$ plane, regular and chaotic regimes are separated by well-defined continuous bands, reflecting the strong sensitivity of the nonlinear response to variations in the resonator stiffness, natural frequency, and electromechanical coupling. By contrast, the $(f_0,I_0)$ map exhibits more fragmented structures elongated along the excitation-frequency direction, indicating that $I_0$ primarily modulates the injected energy rather than the global dynamical organization.

\subsection{Bifurcation Structure and Lyapunov Analysis under Excitation Current and Beam Length Variations}\label{sec4.1}
\begin{figure*}[!t]
    \centering
    \begin{subfigure}[b]{0.45\textwidth}
        \centering
        \includegraphics[width=\textwidth,height=5cm]{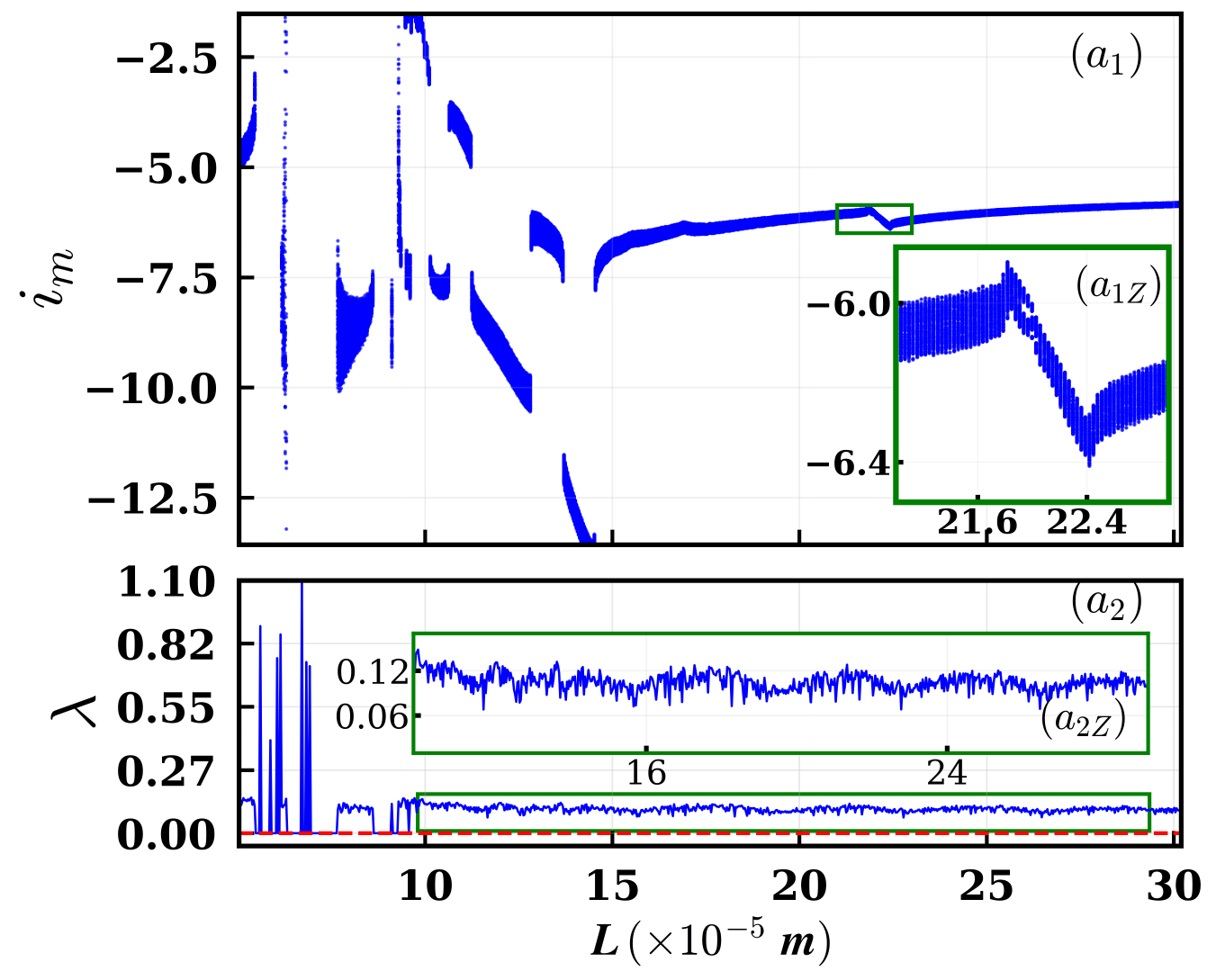}
        \caption{Bifurcation Diagram and Lyapunov Exponent under Beam-Length Variation for $f_0=72.5~\mathrm{kHz}$, $I_0=1~\mathrm{mA}$, $D=25~\mathrm{nm}$, $T=350\,\mathrm{K}$.}
        \label{fig5a}
    \end{subfigure}
    \hfill
    \begin{subfigure}[b]{0.52\textwidth}
        \centering
        \includegraphics[width=\textwidth,height=5cm]{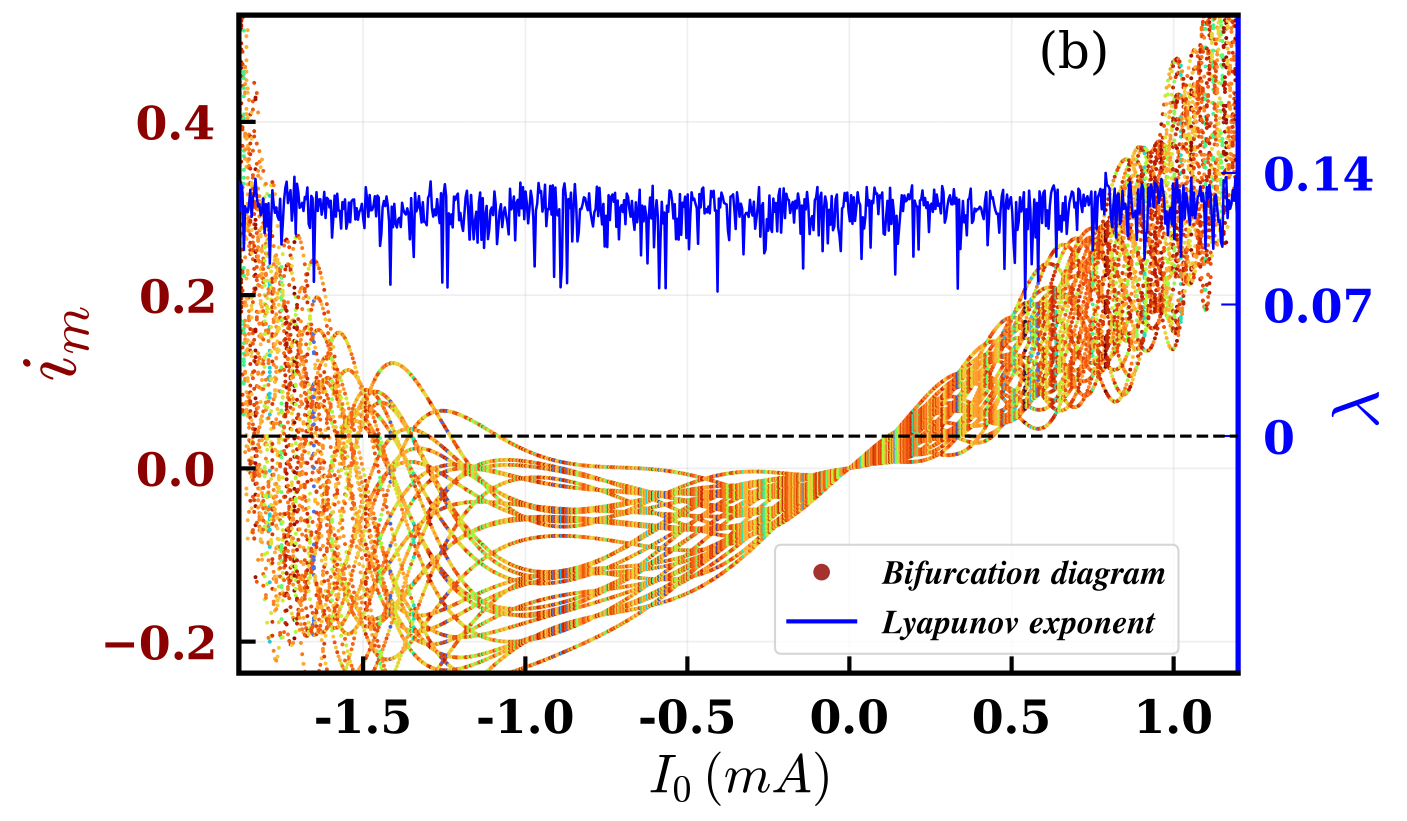}
        \caption{Bifurcation Diagram and Lyapunov Exponent under Excitation-Current Variation for $f_0 = 72.5~\mathrm{kHz}$, $L=15 \times 10^{-5}~\mathrm{m}$, $D=25~\mathrm{nm}$, $T=350\,\mathrm{K}$.}
        \label{fig5b}
    \end{subfigure}
    \caption{\small{Bifurcation structures and Lyapunov exponents of beam-length- and excitation-current-induced dynamical transitions in the thermo-active memristive MEMS. (a) Chaos, bifurcation transitions, and dynamical regularization induced by beam-length variation. (b) Current-induced amplitude modulation and local chaoticity distribution (local variations in chaotic intensity are highlighted by the Turbo colormap).}}
\label{fig5}
\end{figure*}

Guided by the Lyapunov maps, the one-dimensional bifurcation analyses of Fig.~\ref{fig5} are performed over the frequency interval $65\leq f_0\leq80~\mathrm{kHz}$, where the dynamics remain sufficiently rich while limiting frequency-induced reorganizations. Figure~\ref{fig5}(a) is computed for $I_0=1~\mathrm{mA}$ to isolate the influence of the beam length, whereas Fig.~\ref{fig5}(b) is obtained for $L=15\times10^{-5}\,\mathrm{m}$ to emphasize the influence of the excitation current. These parameter selections enable a direct comparison of the respective roles of the structural and electrical control parameters in the nonlinear dynamics of the thermo-active memristive MEMS.

The bifurcation diagrams and corresponding largest Lyapunov exponents in Fig.~\ref{fig5} reveal distinct dynamical responses depending on whether the control parameter is the beam length $L$ or the excitation current $I_0$.
For the beam-length variation (Fig.~\ref{fig5}(a)), the system evolves from a strongly chaotic regime at small $L$ toward progressively more organized oscillatory states through a cascade-like reorganization of the bifurcation structure. Although the largest Lyapunov exponent remains positive, its magnitude decreases significantly, indicating a reduction in chaotic intensity.
By contrast, varying the excitation current $I_0$ (Fig.~\ref{fig5}(b)) primarily modulates the amplitude of the memristor current while preserving the overall chaotic regime, as reflected by the nearly constant positive Lyapunov exponent.
Overall, the results indicate that the beam length mainly governs the global organization and stability of the attractors, whereas the excitation current controls the oscillation amplitude and local chaotic intensity.

\subsection{Bifurcation Structure and Lyapunov Analysis under Frequency and Beam Length Variations}\label{sec4.2}
\begin{figure*}[!t]
    \centering
    \begin{subfigure}[b]{0.49\textwidth}
        \centering
        \includegraphics[width=\textwidth,height=5.3cm]{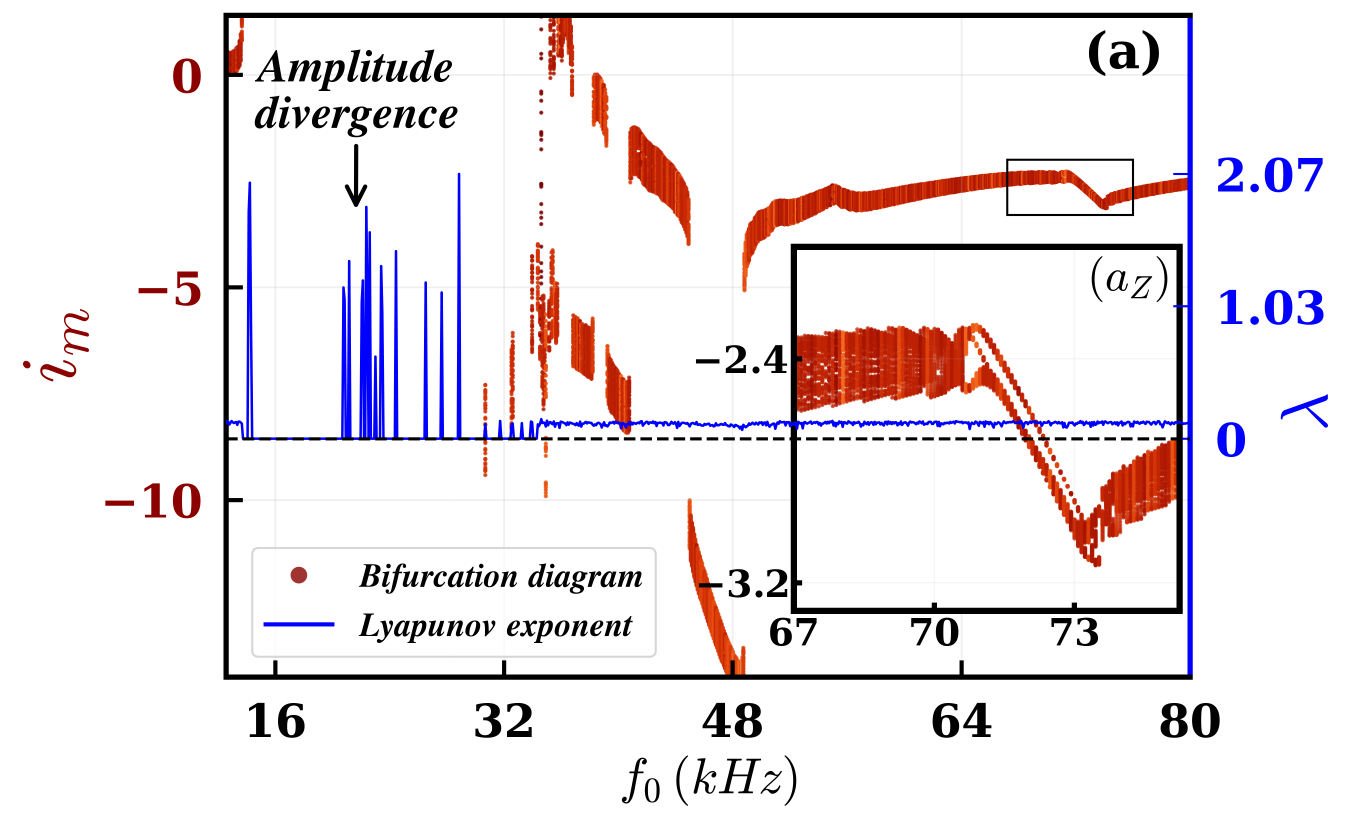}
        \caption{Dynamical transitions under excitation frequency variation for $L=15\times10^{-5}\,\mathrm{m}$.}
        \label{fig6a}
    \end{subfigure}
    \hfill
    \begin{subfigure}[b]{0.48\textwidth}
        \centering
        \includegraphics[width=\textwidth,height=5.3cm]{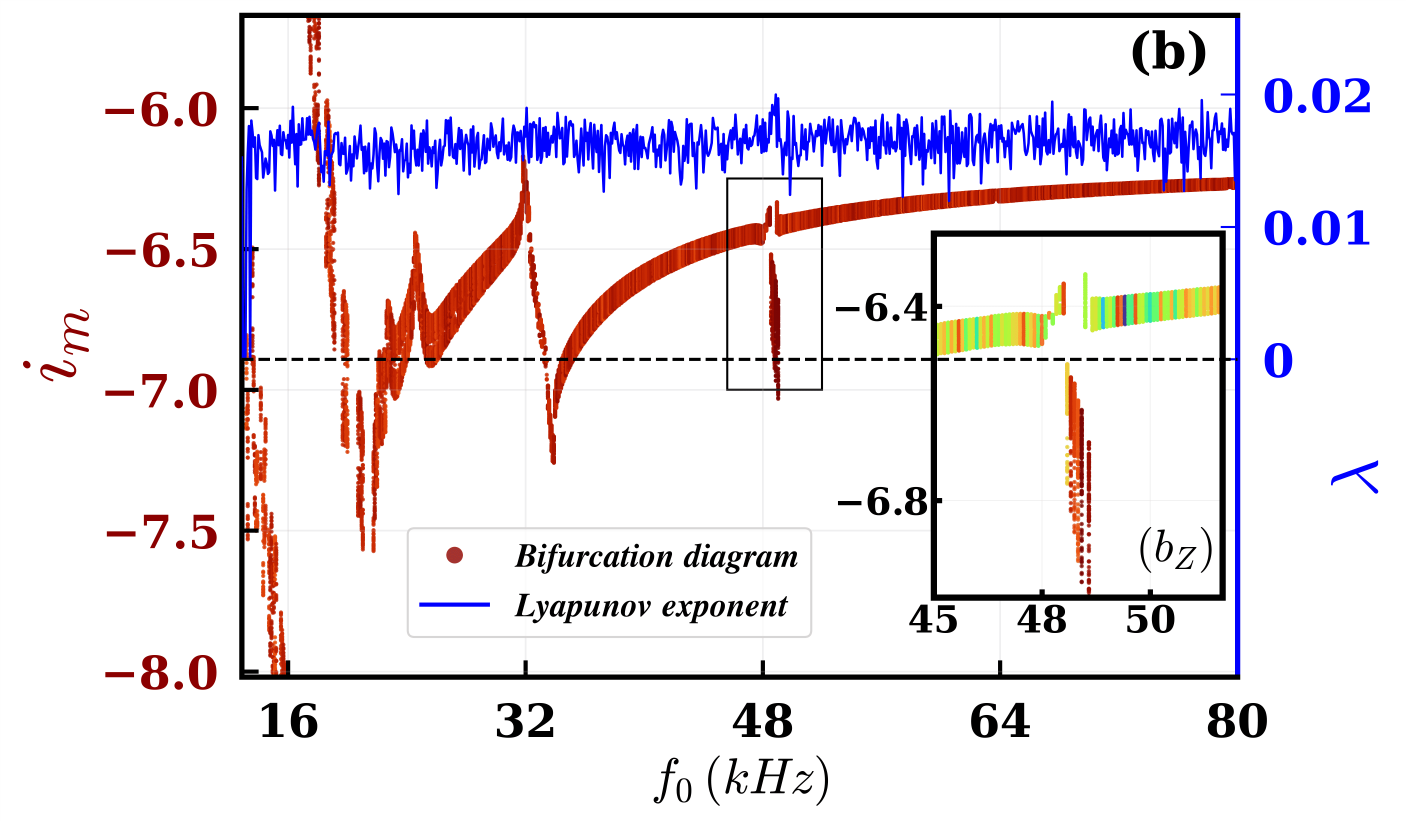}
        \caption{Dynamical transitions under excitation frequency variation for $L=30\times10^{-5}\,\mathrm{m}$.}
        \label{fig6b}
    \end{subfigure}
    \caption{\small{Bifurcation diagrams and corresponding largest Lyapunov exponents as functions of the excitation frequency for two beam lengths, (a): $L=15\times10^{-5}\,\mathrm{m}$ and (b): $L=30\times10^{-5}\,\mathrm{m}$, highlighting the fine organization of attractor branches, and local transitions induced by variations in beam length. The magnified insets $(a_z)$ and $(b_z)$ provide detailed views of selected frequency intervals that are not readily discernible in the full-scale diagrams, (local variations in chaotic intensity are highlighted by the Turbo colormap). \space Parameters: $I_0=1~\mathrm{mA}$, $D=25~\mathrm{nm}$, $T=350\,\mathrm{K}$.}}
\label{fig6}
\end{figure*}

Since the excitation frequency $\omega_0=2\pi f_0$ produces the most pronounced reorganization of the nonlinear dynamical regimes (Figs.~\ref{fig3}(d) and~\ref{fig4}), a more detailed analysis is presented in Fig.~\ref{fig6}, which combines bifurcation diagrams of the memristor current $i_m$ with the corresponding largest Lyapunov exponents for two different beam lengths.

For $L=15\times10^{-5}\,\mathrm{m}$ (Fig.~\ref{fig6}(a)), three frequency regions can be distinguished.

\begin{figure*}[!t]
\centering
  \includegraphics[width=0.9\textwidth]{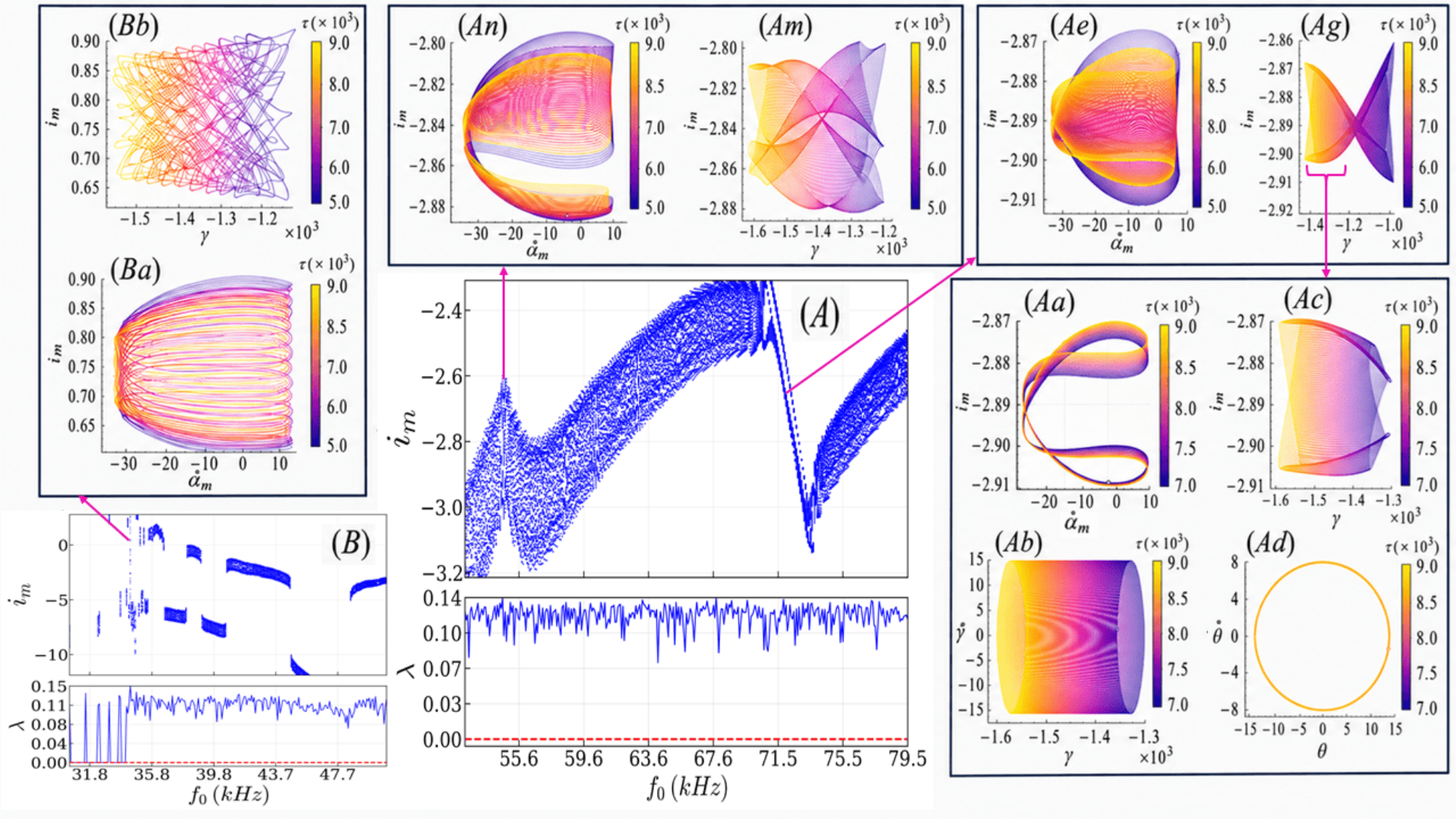}
  \vspace{0.2cm}
\caption{\small{Attractor reconstruction and reorganization in the thermo-active memristive MEMS across two driving-frequency windows. {(A)}, {(B)}: bifurcation diagrams of the memristor current with the largest Lyapunov exponent $\lambda(f_0)$ below (red dashed: $\lambda=0$).
{(Ba)--(Bb)} and {(Aa)--(An)}: attractor projections at the arrowed frequencies, color-coded by $\tau$ ($\times 10^{3}$, $5.0$--$9.0$). Ordered reorganizations in (Aa)--(An),
dispersed interwoven attractors without symmetry in (Ba), (Bb). Parameters:
$T=350$~K, $L=15\times10^{-5}$~m, $D=25$~nm, $I_0=1.2$~mA.}} \label{fig7}
\end{figure*}

Above $50\,\mathrm{kHz}$, the bifurcation structure is more organized, and the attractor branches cluster into narrower families, indicating reduced chaotic intensity despite the persistence of positive Lyapunov exponents. The inset in Fig.~\ref{fig6}$(a_z)$ further reveals fine-scale attractor reorganizations and localized variations in chaotic intensity (see Fig.~\ref{fig7}(A), $(Aa)$--$(An)$). Region~A (Fig.~\ref{fig7}) exhibits persistent deterministic chaos associated with pronounced topological reorganizations as the excitation frequency varies, leading to extended folded manifolds and compact bi-lobed attractors (see Fig.~\ref{fig7}, $(An)$, $(Am)$ for $f_0=52.8~\mathrm{kHz}$ and $(Ae)$, $(Ag)$ for $f_0=72.55~\mathrm{kHz}$). Detailed phase-space projections (Fig.~\ref{fig7}(A), $(Aa)$--$(Ad)$) reveal that these attractors possess a hierarchical geometrical organization for $f_0 = 72.5~\mathrm{kHz}$, where the mechanical subsystem remains nearly harmonic while the memristive variables undergo strong nonlinear folding and coupling. By contrast, from $32$ to $50\,\mathrm{kHz}$ (Fig.~\ref{fig6}(a)), repeated branch splitting and merging reveal successive attractor reorganizations while the Lyapunov exponent remains predominantly positive. Between approximately $15$ and $30\,\mathrm{kHz}$ (Fig.~\ref{fig6}(a)), the broad spreading of the bifurcation branches and pronounced Lyapunov peaks indicate intense chaotic activity characterized by fragmented bifurcation branches and larger Lyapunov fluctuations, giving rise to highly dispersed attractors with extensive orbit interweaving and no apparent geometrical symmetry (see Fig.~\ref{fig7}(B), $(Ba)$ and $(Bb)$ for $f_0 = 35~\mathrm{kHz}$). These observations reveal that the excitation frequency continuously reshapes the attractor topology and chaotic intensity, while the beam length acts as an effective control parameter governing the complexity of the thermo-active memristive MEMS.

\begin{figure*}[!t]
\centering
  \includegraphics[width=0.88\textwidth]{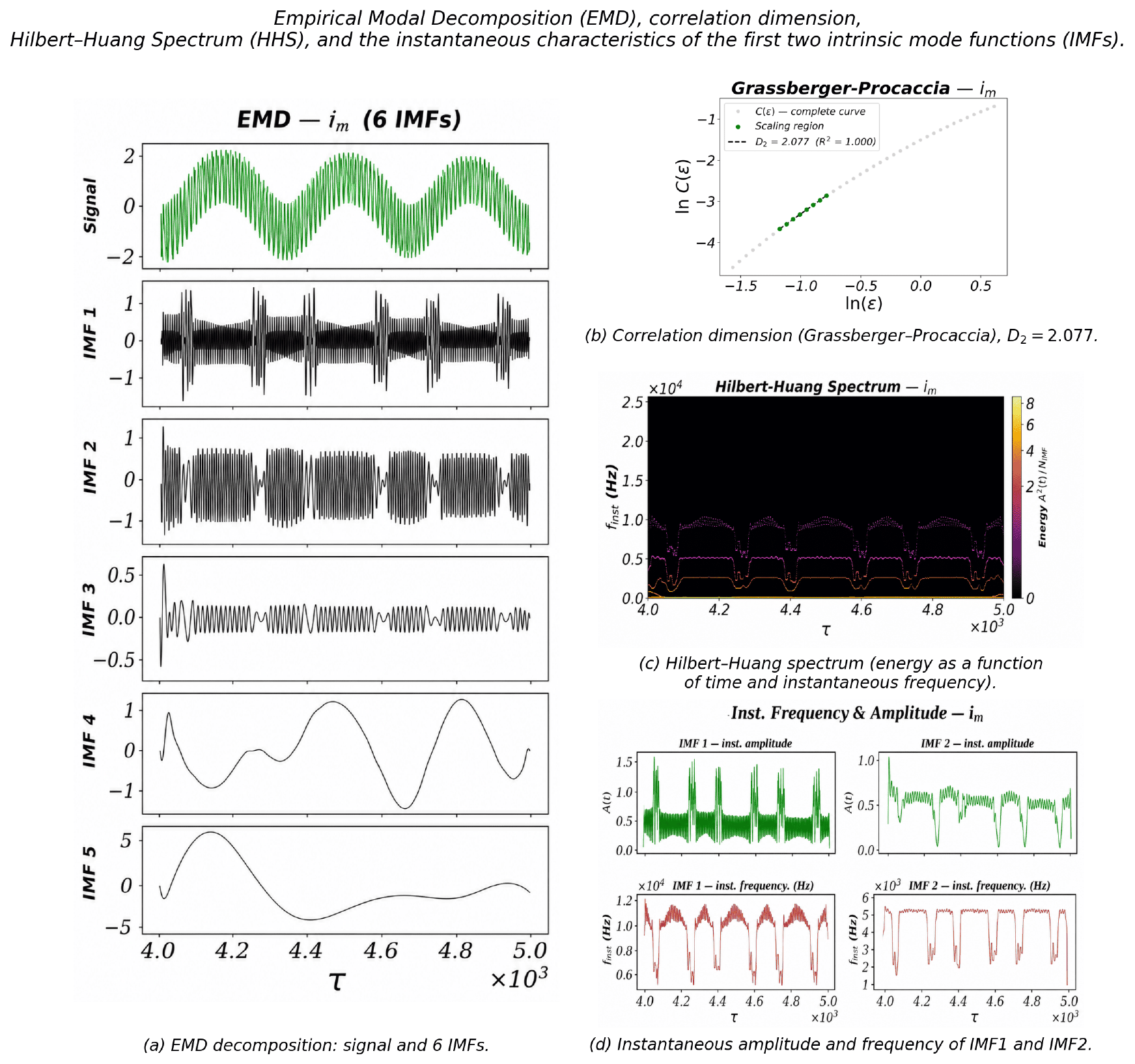}
  \vspace{0.2cm}
  \caption{\small{Detailed multiscale characterization of the deterministic chaotic window highlighted in the zoomed bifurcation diagram of Fig.~\ref{fig6}(b$_z$) at $f_0 \approx 49\,\mathrm{kHz}$. \space Parameters: $I_0=1~\mathrm{mA}$, $L=30\times10^{-5}\,\mathrm{m}$, $D=25~\mathrm{nm}$, $T=350\,\mathrm{K}$.}}\label{fig8}
\end{figure*}

A markedly different behavior is observed for $L=30\times10^{-5}\,\mathrm{m}$ (Fig.~\ref{fig6}(b)). The bifurcation branches remain comparatively compact over the entire frequency range, indicating that increasing the beam length reduces the amplitude variability without suppressing the complex nonlinear dynamics. Although the largest Lyapunov exponent remains positive, its fluctuations are weaker, reflecting a more structured chaotic regime. The inset in Fig.~\ref{fig6}$(b_z)$ reveals local attractor reorganizations accompanied by localized peaks of the Lyapunov exponent, suggesting repeated transitions between weakly chaotic and quasi-periodic states. The nonlinear analyses presented in Fig.~\ref{fig8} focus on the operating frequency $f_0\approx49~\mathrm{kHz}$, corresponding to the enlarged bifurcation window of Fig.~\ref{fig6}(b$_z$), where a local increase in the largest Lyapunov exponent and a sudden bifurcation reorganization indicate enhanced dynamical complexity. The Empirical Mode Decomposition (EMD) with the Intrinsic Mode Function (IMF) of the memristive current $i_m$ (Fig.~\ref{fig8}(a),\,(d)), and the Hilbert--Huang analyses (Fig.~\ref{fig8}(c)) reveal that this chaotic behavior results from nonlinear interactions between multiple intrinsic oscillatory modes, leading to pronounced amplitude modulation and non-stationary energy redistribution. The Grassberger--Procaccia algorithm (Fig.~\ref{fig8}(b)) yields a correlation dimension of $D_2=2.077$, indicating the evolution on a low-dimensional deterministic chaotic attractor, consistent with the positive Lyapunov exponent. 

Together, these complementary diagnostics indicate that $f_0\approx49~\mathrm{kHz}$ corresponds to a critical operating condition where nonlinear mode interactions drive the transition to deterministic chaos in the thermo-controlled memristive resonator system.
The combined interpretation of the bifurcation diagrams and the corresponding Lyapunov exponents in Fig.~\ref{fig6}, together with the reconstructed attractors presented in Figs.~\ref{fig7} and~\ref{fig8}, reveals that variations in the excitation frequency $f_0$ and the beam length $L$ progressively reorganize both the global topology of the attractors and the local distribution of chaotic activity.

\subsection{Frequency Ratio as an Organizing Parameter of the Nonlinear Dynamics}\label{sec4.3}
A strong similarity is observed between the bifurcation diagrams
obtained by varying the excitation frequency $f_0$ (Fig.~\ref{fig6}(a))
and the beam length $L$ (Fig.~\ref{fig5}(a)). Both exhibit comparable
chaotic organizations with repeated attractor reorganizations and
predominantly positive Lyapunov exponents, indicating that these two
physically distinct control parameters drive the system through a
common underlying mechanism. This correspondence follows from the
dependence of the microbeam natural frequency on its geometry.
According to Euler--Bernoulli beam theory~\cite{Ekinci1},
\begin{equation}
\omega_b \simeq \frac{h}{L^2}\sqrt{\frac{E}{\rho}},
\label{eq:wb}
\end{equation}
so that increasing $L$ decreases $\omega_b$ and modifies the frequency
ratio
\begin{equation}
r_{\omega} = \frac{\omega_0}{\omega_b}.
\end{equation}
Varying either $L$ or $f_0$ therefore shifts the same frequency
detuning, explaining the close correspondence between the two
bifurcation scenarios. This interpretation is consistent with the
Lyapunov map of Fig.~\ref{fig4}(b), where the dynamical regimes are
organized along curved bands in the $(f_0,L)$ plane, suggesting that
the global dynamics are governed by $r_\omega$, which couples the
effects of $L$ (through $\omega_b$) and $f_0$ (through $\omega_0$).
It should be noted that $L$ also enters the governing equations
independently of $\omega_b$, most notably through the beam resistance
$r_b = \mu L/A$. However, being a purely resistive term, $r_b$ can only
affect current amplitude and dissipation; it cannot, by itself, shift
the characteristic oscillation frequency of the system.

\begin{figure*}[!t]
\centering
\includegraphics[width=0.85\textwidth,height=2cm,trim=0 10 0 0.5,clip]{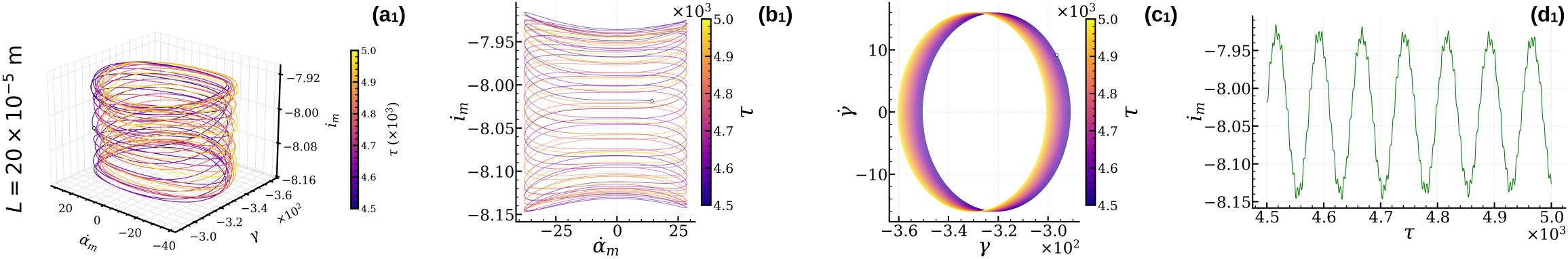}\\[0.2cm]
\includegraphics[width=0.85\textwidth,height=2cm]{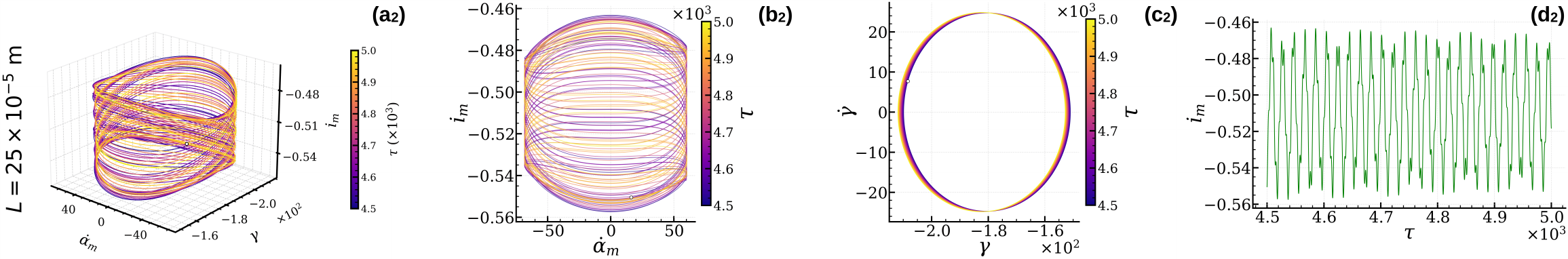}\\[0.2cm]
\includegraphics[width=0.85\textwidth,height=2cm]{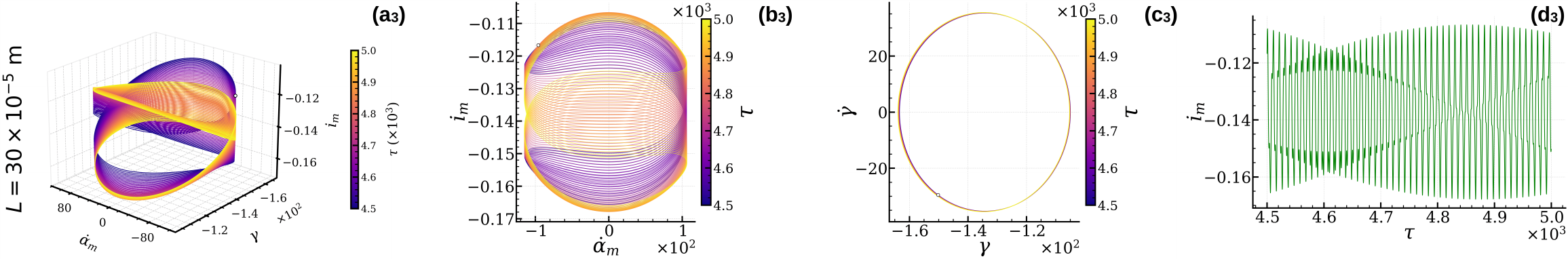}
  \vspace{0.2cm}
  \caption{\small{Evolution of reconstructed attractors (a$_1$--a$_3$),
  electro-memristive phase-space projections (b$_1$--b$_3$),
  resonator-current phase portraits (c$_1$--c$_3$), and memristive
  current time responses (d$_1$--d$_3$) with increasing beam
  length. \space Parameters: $I_0=1.2\,\mathrm{mA}$, $f_0=38.85~\mathrm{kHz}$, $D=15~\mathrm{nm}$, $T=350\,\mathrm{K}$.}} \label{fig9}
\end{figure*}

Decisive confirmation is provided by the temporal responses of
Fig.~\ref{fig9} (panels~$(d_1)$--$(d_3)$): the oscillation frequency of
$i_m$ increases monotonically with $L$ despite the excitation frequency
$f_0$ being held strictly constant. Since a purely resistive term such
as $r_b$ cannot produce such a frequency shift, this observation can
only result from a genuine variation of the system's resonance
condition; that is, from the variation of $\omega_b$ itself. This
frequency-domain signature therefore identifies $r_\omega$, rather than
the secondary geometric dependence of $L$ through $r_b$, as the
physical mechanism governing the observed dynamics.

This organizing role is further reflected in the attractor geometry.
For a fixed excitation frequency, increasing $L$ and hence
$r_\omega$ progressively regularizes the dynamics: the attractor
evolves from a densely intertwined structure at $L = 20\times10^{-5}$~m
to a smoother folded manifold at $L = 30\times10^{-5}$~m
(panels~$(a_1)$--$(a_3)$), while the mechanical portraits approach
nearly elliptical closed orbits indicative of an almost harmonic
response (panels~$(c_1)$--$(c_3)$).

This broad chaotic domain, jointly controlled by the beam length
(Figs.~\ref{fig6} and~\ref{fig9}) and the excitation frequency
(Figs.~\ref{fig6} and~\ref{fig7}), could offer a compact micro-scale
source of deterministic chaos for chaos-based secure communication and
random signal generation, with the frequency ratio $r_\omega$ serving
as a practical design rule for placing the operating point in either
the regular or chaotic regions.

\section{Thermal Optimization of Memristive Hysteresis and Topological Reconfiguration}\label{sec5}
\begin{figure*}[!t]
    \centering
    \begin{subfigure}[b]{0.47\textwidth}
        \centering
        \includegraphics[width=\textwidth,height=4.5cm]{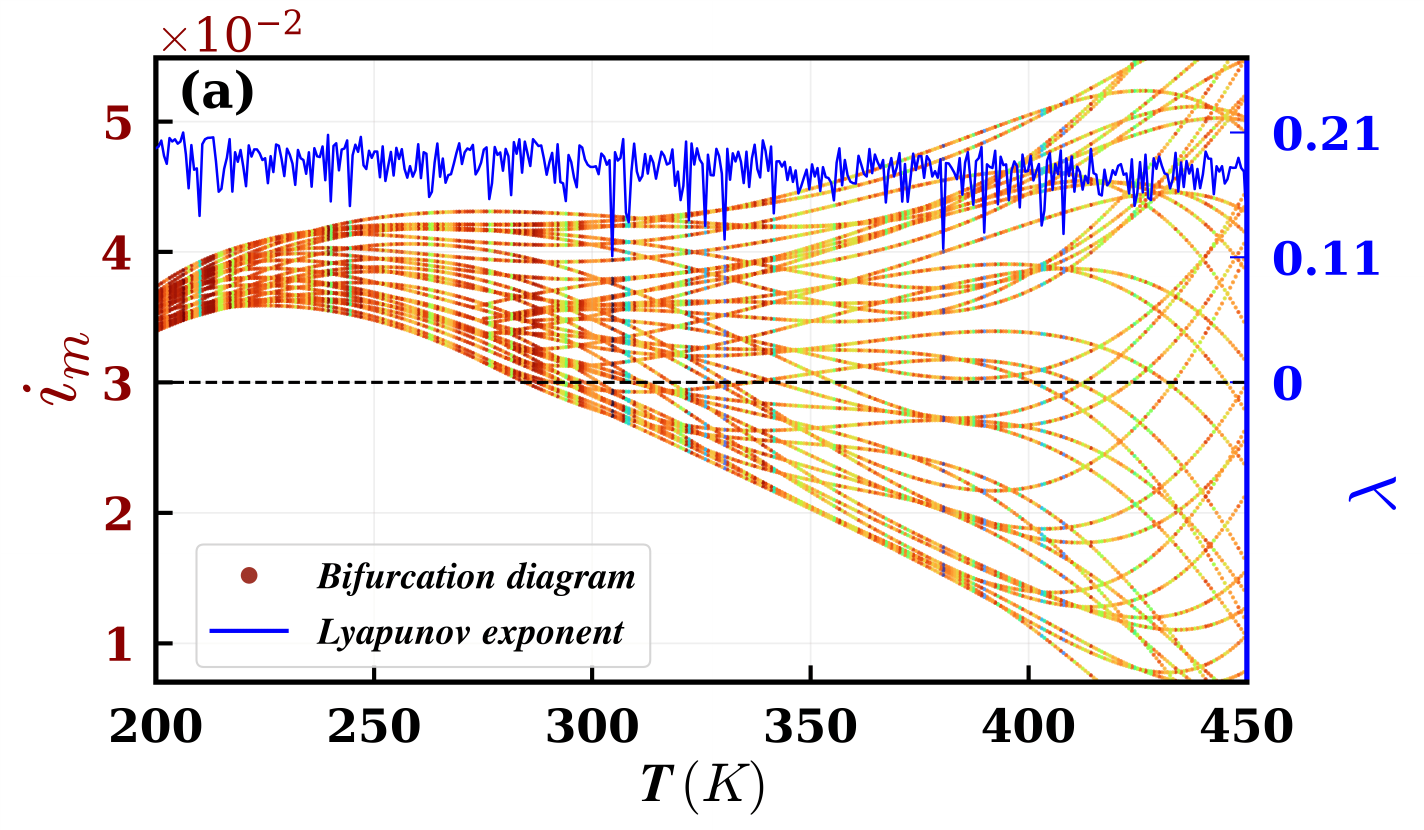}
        \caption{Bifurcation diagrams and corresponding largest Lyapunov exponents as functions of the Temperature.}
        \label{fig12a}
    \end{subfigure}
    \hfill
    \begin{subfigure}[b]{0.47\textwidth}
        \centering
        \includegraphics[width=\textwidth,height=4.1cm]{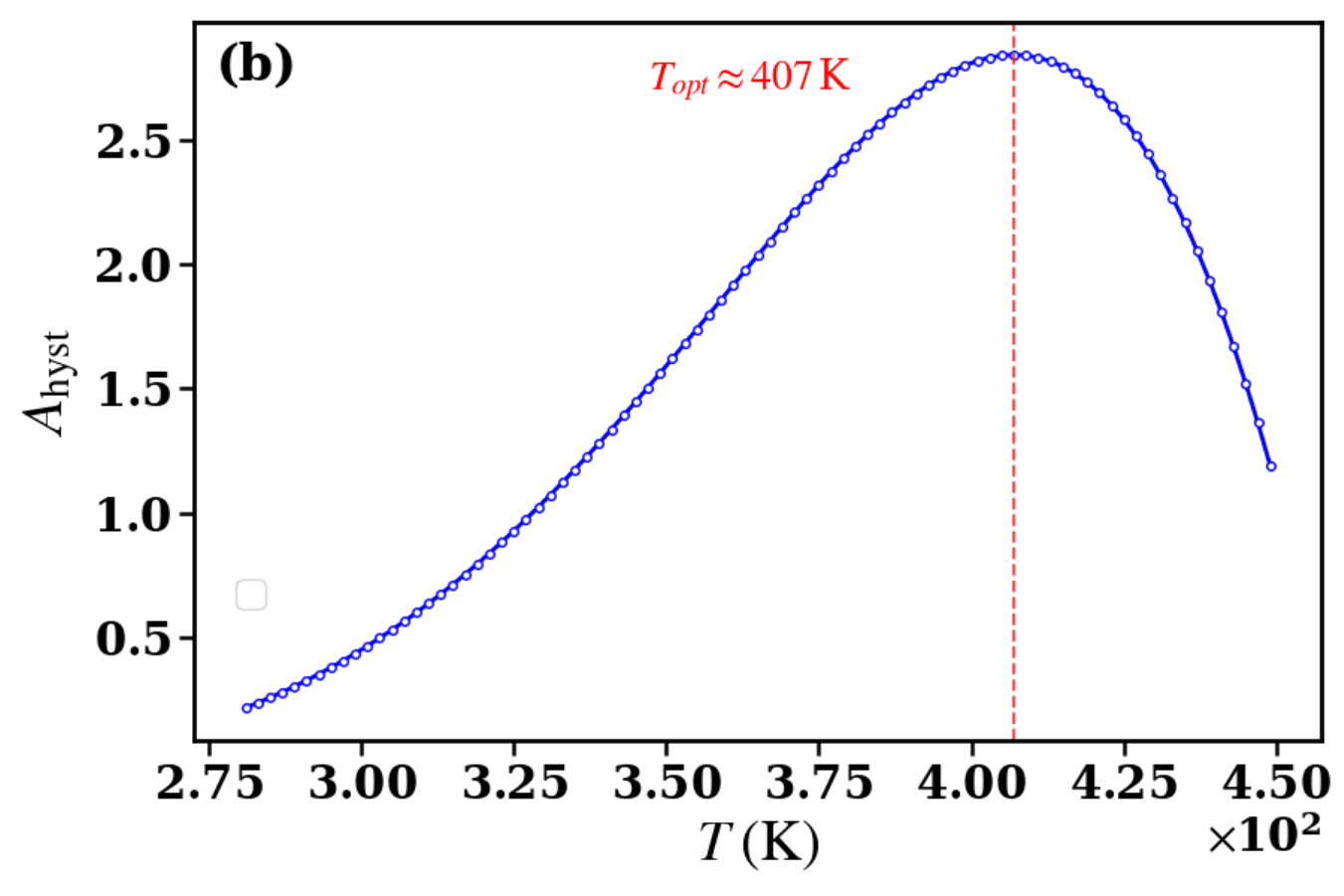}
        \caption{Memristive hysteresis area with temperature and identification of the optimal thermo-memristive operating point.}
        \label{fig12b}
    \end{subfigure}
    \caption{\small{(a): Bifurcation diagrams and corresponding largest Lyapunov exponents as functions of the temperature, highlighting the organization of attractor branches, and local transitions induced by variations in temperature, (local variations in chaotic intensity are highlighted by the Turbo colormap). (b): Quantitative evolution of the enclosed hysteresis area with temperature and identification of the optimal thermo-memristive operating point. Parameters: $L=30\times10^{-5}\,\mathrm{m}$, $D=25\,\mathrm{nm}$, $I_0=1\,\mathrm{mA}$, $f_0 = 38.85\,\mathrm{kHz}$.}}
\label{fig12}
\end{figure*}

Fig.~\ref{fig12} characterizes the influence of temperature on the
nonlinear electromechanical dynamics through bifurcation analysis,
largest Lyapunov exponent (Fig.~\ref{fig12a}), and hysteresis quantification (Fig.~\ref{fig12b}).

At low temperatures ($T \lesssim 280$~K), the attractor remains
spatially confined and is predominantly characterized by red-colored
branches, indicating a strongly nonlinear dynamical regime. As the
temperature increases toward $T \approx 280$--$400$~K, the attractor
branches progressively expand and reorganize while the red coloration
gradually diminishes, reflecting a weakening of the nonlinear response.
Beyond $T > 400$~K, the branches progressively converge without
completely merging into a single branch. The temperature, therefore, acts as
a topological control parameter driving the attractor from a spatially
confined and strongly nonlinear state at low temperatures toward a more
spatially extended, yet dynamically weakened, configuration at elevated
temperatures.

These thermally induced reorganizations are directly reflected in the hysteresis area $A_{\mathrm{hyst}}$. As shown in Fig.~\ref{fig12b}, $A_{\mathrm{hyst}}$ increases up to a maximum at $T_{\mathrm{opt}}\approx 407\,\mathrm{K}$ and decreases thereafter, revealing an optimal thermo-memristive operating point. At this temperature, nonlinear electrical conduction, thermally activated ionic mobility, and electromechanical coupling achieve their most favorable balance. Since this balance depends on the device geometry and electrical forcing, $T_{\mathrm{opt}}$ is expected to vary with $L$, $D$, and $I_0$, indicating an interior optimum rather than a universal temperature. The optimal temperature, $T_{\mathrm{opt}} \approx 407\,\mathrm{K}$ ($\approx 134\,^{\circ}\mathrm{C}$), lies well below the thermal stability limits of silicon and TiO$_2$-based devices. Rather than representing a universal optimum, it reflects a design trade-off in which the ionic mobility maximizing $A_{\mathrm{hyst}}$ also promotes faster state relaxation. Since $T_{\mathrm{opt}}$ depends on $L$, $D$, and $I_0$, it should be regarded as a tunable design parameter rather than a fixed constraint.

\begin{figure*}[!t]
\centering
  \includegraphics[width=0.78\textwidth,height=7.5cm]{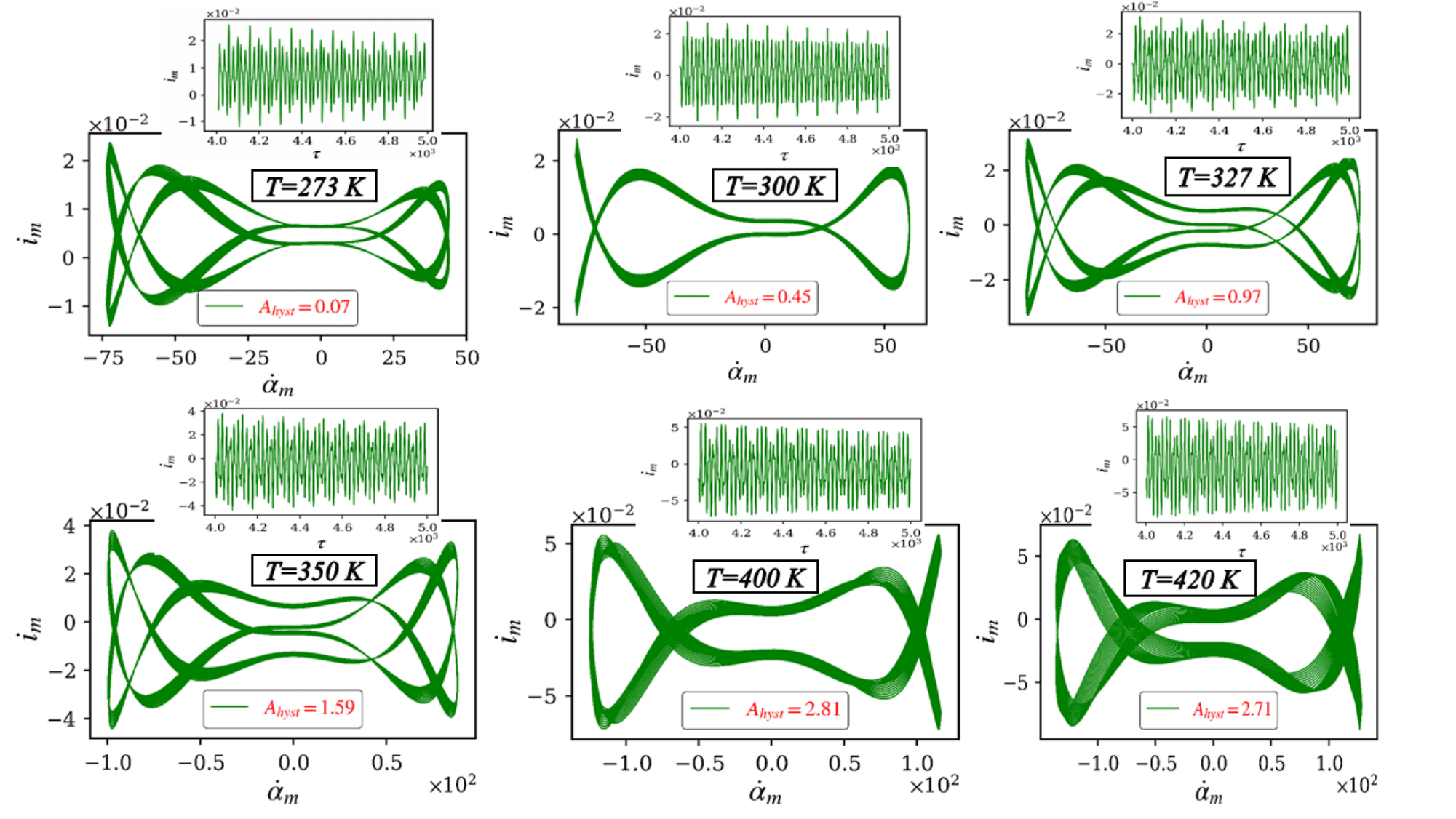}
  \caption{\small{Thermally induced topological reconfiguration of electro-memristive hysteresis loops and quantitative evolution of the enclosed hysteresis area $A_{\mathrm{hyst}}$. Parameters: $L=30\times10^{-5}\,\mathrm{m}$, $D=25\,\mathrm{nm}$, $I_0=1\,\mathrm{mA}$, $f_0 = 38.85\,\mathrm{kHz}$.}} \label{fig13}
\end{figure*}

Figure~\ref{fig13} illustrates the thermally induced evolution of the electro-memristive hysteresis loops together with the corresponding temporal responses and the enclosed hysteresis area $A_{\mathrm{hyst}}$. As the temperature increases from $T=273\,\mathrm{K}$ to $T=420\,\mathrm{K}$, the hysteresis loops undergo a progressive topological reconfiguration while preserving their characteristic pinched geometry. Rather than producing an abrupt transition between distinct dynamical regimes, thermal activation continuously reshapes the geometry of the hysteresis loops through successive modifications of their width, curvature, and enclosed area, revealing the gradual reorganization of the underlying nonlinear electro-memristive dynamics.

Taken together, these results (Figs.~\ref{fig12} and \ref{fig13}) show
that temperature is an efficient, non-invasive, and continuously tunable
control parameter for adaptive memristive MEMS. The underlying
thermo-memristive coupling can be summarized by the sequence
$
T \to \sigma(T) \to M(w,T)
\to i_m(t) \to w(t),
$
through which temperature regulates the nonlinear dynamics by successively
modifying the electrical conductivity, memristance, current, and internal
state of the device. For $L=30\times10^{-5}\,\mathrm{m}$, the operating
window ($280$--$407\,\mathrm{K}$) simultaneously optimizes the attractor
complexity and hysteretic memory. Together with geometry at the design stage
and electrical excitation during operation, temperature forms a three-level
control strategy enabling temperature-programmable neuromorphic synapses,
resonant thermal sensors, and intelligent MEMS integrating sensing, memory,
and nonlinear signal processing within a single device.

\section{Conclusion}\label{sec:conclusion}\label{sec6}
This work highlights a comprehensive thermo-electro-mechanical framework for investigating the nonlinear dynamics of a memristive resonator composed of a doubly clamped Euler--Bernoulli microbeam, an $RLC$ resonant circuit, and a TiO$_2$ memristor with temperature-dependent ionic transport. By coupling the thermally activated evolution of the memristor internal state with the thermo-geometrical properties of the mechanical subsystem, the proposed framework provides a unified description of the coupled nonlinear dynamics. The nonlinear analysis reveals that the coupled dynamics are governed by the interplay between geometry, electrical excitation, and thermal activation, leading to four major findings. First, Lyapunov--Benettin parameter space maps and reconstructed attractors show that the dynamics are predominantly quasi-periodic or chaotic, with thermal loading, beam geometry, and electrical excitation continuously reorganizing the distribution of nonlinear regimes across the parameter space. Second, complementary nonlinear diagnostics (Hilbert--Huang spectra,
Poincar\'e sections, and Grassberger--Procaccia correlation dimensions)
identify the thermo-memristive subsystem as the primary source of the
nonlinear complexity, which is transmitted to the vibrating beam through the
electromechanical coupling; under fixed operating conditions, the asymptotic
regime is found to be initial-state dependent. Third, the beam length and the excitation frequency govern the global bifurcation organization through the common frequency ratio $r_\omega=\omega_0/\omega_b$, whereas the excitation current mainly controls the oscillation amplitude and local chaotic intensity. Finally, temperature continuously reorganizes the bifurcation structure, the nonlinear attractors, and the electro-memristive hysteresis through the thermo-memristive coupling mechanism, while providing an efficient means of tuning the memristive memory and defining an optimal operating region near $T\approx407$~K. These results open several promising application perspectives, including
nonvolatile mechanical memories and reconfigurable logic enabled by the
observed initial-state-dependent regime selection, temperature-programmable
synaptic elements for neuromorphic architectures, resonant thermal sensing
based on the thermal sensitivity of the hysteresis area $A_{\mathrm{hyst}}(T)$, deterministic chaos generators for secure communications and random signal generation, and nonlinear signal processing, where the frequency ratio
$r_\omega$ provides a practical design criterion for selecting regular or
chaotic operating regimes.

Although the present investigation is entirely numerical, each constitutive component of the proposed model relies on experimentally established physical laws, including the temperature dependence of $R_{\mathrm{on}}/R_{\mathrm{off}}$, the Arora mobility model, and the Mott/Efros--Shklovskii hopping conduction mechanisms discussed in Sec.~\ref{sec2}. The initial-state dependence reported here further suggests an underlying multistability of the coupled system; a systematic mapping of the associated basins of attraction, together with the experimental validation of the proposed thermo-active memristive MEMS, is left for future work.
\section*{Appendix}
\begin{appendices}
\section{Equations and Parameters Expression}\label{Sec7}
\renewcommand{\theequation}{\thesection.\arabic{equation}}
\setcounter{equation}{0}
\begin{table*}[!t]
\centering
\caption{Mathematical expressions of the parametric coefficients of the system, Eq.~(\ref{eq15})}
\label{System_Param}
\footnotesize
\begin{tabular}{>{\centering\arraybackslash}p{0.09\textwidth}
                p{0.34\textwidth}
                @{\hspace{1.4em}}>{\centering\arraybackslash}p{0.09\textwidth}
                p{0.34\textwidth}}
\toprule
\textbf{Param.} & \hspace{1.2cm}\textbf{Expression} &\textbf{Param.} &\hspace{1.2cm} \textbf{Expression} \\
\midrule
$\beta_1$ &\hspace{1.2cm} $\dfrac{\lambda}{\omega_1\rho A} + \dfrac{16 B^2 L}{15\,\omega_1\mu\rho}$
 & \hspace{1.5cm}$\omega_0^{*}$ &\hspace{1.2cm} $\dfrac{\omega_0}{\omega_1}$ \\
 \\
$\delta_{r_2}$ &\hspace{0.5cm} $\left(\dfrac{r_p + r_0}{L_0} + \dfrac{16 B^2 L}{15\mu\rho}\right)\omega_1 Q_0$
 &\hspace{1cm} $I_0^{*}$ & \hspace{1.2cm}$I_0\,\omega_0$ \\
 \\
$\Delta_{R_m}$ & \hspace{1.2cm}$\mu_v\dfrac{(R_{\mathrm{off}}-R_{\mathrm{on}})R_{\mathrm{on}}}{D^2\omega_1^2\phi_0}$
 & $\vartheta_0$ &\hspace{1.2cm} $\omega_1\phi_0$ \\
 \\
$\sigma_5$ & \hspace{1.2cm}$\dfrac{16 B\,E I_y \xi^4 T_0}{15\,\mu\rho}$
 & $\vartheta_2$ & \hspace{1.2cm}$\dfrac{\omega_1\phi_0}{L_0}$ \\
 \\
$\beta_2$ & \hspace{1.2cm}$\dfrac{E I_y \xi^4}{\rho\,\omega_1^2 A}$
 & $R_0^{*}$ & \hspace{1.2cm}$\dfrac{R_{\mathrm{off}}}{R_{\mathrm{on}}}$ \\
 \\
$\delta_{r_1}$ &\hspace{1.2cm} $\dfrac{r_p + r_0}{L_0\,\omega_1}$
 & $\Delta_{R_{\mathrm{on}}}$ & \hspace{1.2cm}$\mu_v\dfrac{R_{\mathrm{off}}-R_{\mathrm{on}}}{D^2}$ \\
 \\
$\vartheta_1$ &\hspace{1.2cm} $\dfrac{\phi_0}{L_0\,Q_0\,\omega_1}$
 & $\omega_{C_0}^2$ &\hspace{1.2cm} $\dfrac{1}{L_0\,C_0}$ \\
 \\
$\beta_3$ &\hspace{1.2cm} $\dfrac{B\,L\,Q_0}{\rho\,\omega_1 A\,T_0}$
 & $\Omega_{C_0}^2$ & \hspace{1.2cm}$\dfrac{1}{\omega_1^2 L_0\,C_0}$ \\
 \\
$\sigma_2$ & \hspace{1.2cm}$\dfrac{16 B A T_0}{15\mu}$
 & $\sigma_4$ &\hspace{1.2cm} $\dfrac{Q_0}{L_0 C_0}$ \\
 \\
$\sigma_0$ & \hspace{1.2cm}$\dfrac{16 B A\,\omega_1 T_0}{15\mu}$
 & $\sigma_1$ & \hspace{1.2cm}$\dfrac{R_{\mathrm{on}}}{\omega_1^2\phi_0}$ \\
 \\
$\sigma_3$ &
 $\dfrac{16 B L\,\omega_1 T_0}{15}\left(\dfrac{1}{L_0} + \dfrac{\lambda}{\mu\rho L}
 + \dfrac{16 B^2 A}{15\mu^2\rho}\right)$ & $\eta$ &\hspace{1.2cm} $\dfrac{16 B L}{15 L_0 \omega_1}$ \\
 \\
\bottomrule
\end{tabular}
\end{table*}

The dynamics of the microbeams can be described by the partial differential equation Eq.~(\ref{eq::mod1}), derived from the classical Euler--Bernoulli beam theory~\cite{Timoshenko1955,Nayfeh2004,Pelesko2002,Senturia2001,Younis2011}.
\begin{equation}
    EI_y \dfrac{\partial^4 u(z,t)}{\partial z^4}  + \rho A \dfrac{\partial^2 u(z,t)}{\partial t^2} +  \lambda \dfrac{\partial u(z,t)}{\partial t} + NLT = f(t),
    \label{eq::mod1}
\end{equation}
where the nonlinearity term is assumed to be negligible, i.e., $NLT = 0$.
The actuating force, $f(t)$ is a Lorentz force due to the current $i_b$ through the beam and is expressed as~\cite{Koudaf,Koudaf2,Koudaf3}:
 \begin{equation}
 \begin{cases}
     f(t) = B \, L \,i_b  =
B \, L \,\left( \dfrac{dq}{dt} + \dfrac{e_b}{r_b}\right)\\
\dfrac{e_b}{r_b}
=
\dfrac{-2\,BA}{\mu L}\,
\displaystyle\int_{0}^{L}
\frac{\partial u(z,t)}{\partial t}\,dz.
 \end{cases}
\label{eq::mod2}
 \end{equation}
with $\vec{B} \perp L\vec{l}$ and

By substituting Eq.~(\ref{eq::mod2}) into Eq.~(\ref{eq::mod1}), we obtain the general model expressed in Eq.~(\ref{eq::eqt1}).
\begin{multline}\label{eq::eqt1}
EI_y \frac{\partial^4 u(z,t)}{\partial z^4}  + \rho A \frac{\partial^2 u(z,t)}{\partial t^2} +
\lambda \frac{\partial u(z,t)}{\partial t} =\\
B \, L \,\left[\frac{dq}{dt} -  \frac{2\,B\,A}{\mu\,L}\int_{0}^{L} \frac{\partial u(z,t)}{\partial t} dz\right].
\end{multline}
The shape of the modes must satisfy the differential equation and boundary conditions \cite{Han1999}.
\begin{equation}
    \frac{\partial^4
	u(z,t)}{\partial z^4} + \frac{\rho A}{E\,I_y}\frac{\partial^2
	u(z,t)}{\partial t^2} = 0 .
\end{equation}
In our case, we have $ u(0,t) =u(L,t) = 0$ (boundary conditions), and the set of eigenfunctions $Z_n(z)$ is written:
\begin{equation}
   Z_n(z) = a_n(\cos \xi_n\,z - \cosh \xi_n\,z) +
b_n(\sin\xi_n\,z - \sinh\xi_n\,z)
\end{equation}
with \,$\xi_n$\, the
solution of the transcendental equation:
\begin{equation}
 \cos \xi_n\,L\,\cosh \xi_n\,L - 1 = 0 .
\end{equation}
As we focus on the study of the fundamental state, we shall take it for the rest $n=1$.

In addition, for this study we use the normalization made by N. Lobontiu \cite{Lobontiu1,Lobontiu2}:
$Z_1(z) = Z_1(\frac{L}{2}) = 1$
and
$ \int_{0}^{L}Z_n(z)dz = 0.5333\, L = \dfrac{8}{15} L $.
Integrating all these transformations into Eq.~(\ref{eq::eqt1}), we have Eq.~(\ref{eq8}).

Applying Kirchhoff's voltage law (KVL) $V_m - U_{\mathrm{beam}} - U_{C_0} - U_{L_0} = 0$, and by differentiating with respect to time, we get the memristor flux equation:
\begin{figure*}[!h]
\centering
\footnotesize
\begin{equation}\label{eq12}
\begin{split}
\ddot{\phi}_m(t) = &- \mu_v \dfrac{(R_{\mathrm{off}} - R_{\mathrm{on}}) R_{\mathrm{on}}}{D^2}
        \Big\{I_0\cos(\omega_0 t) - \dot{q}(t) +
        \dfrac{16 B A}{15 \mu}\dot{u}(t)\Big\}^2 \\
        &+ R_{\mathrm{on}}\Bigg\{  \dfrac{R_{\mathrm{off}}}{R_{\mathrm{on}}} - \mu_v \dfrac{(R_{\mathrm{off}} - R_{\mathrm{on}})}{D^2}
        \Big\{  \dfrac{I_0}{\omega_0}\sin(\omega_0 t) -
        q(t) + \dfrac{16 B A}{15 \mu}u(t)\Big\} \Bigg\} \\
        &\times \Bigg\{-I_0\,\omega_0\sin(\omega_0 t)
    - \dfrac{1}{L_0}\dot{\phi}_m(t)
    + \Big\{\dfrac{r_p + r_0}{L_0} +
    \dfrac{16\,B^2\,L}{15\,\mu\,\rho}\Big\}\dot{q}(t) - \dfrac{16 B L}{15}\Big\{\dfrac{1}{L_0} + \dfrac{\lambda}{\mu\,\rho\,L} +
    \dfrac{16\,B^2\,A}{15 \mu^2\,\rho}\Big\}\dot{u}(t) \\
    &\qquad + \dfrac{1}{L_0C_0}q(t)
    - \dfrac{16\,B\,E I_y {\xi_n}^4}{15\,\mu\,\rho}u(t)\Bigg\}
\end{split}
\end{equation}
\end{figure*}
The various coefficients of the dynamic system in Eq.~(\ref{eq12}), obtained through mathematical modeling, are listed in Table~\ref{System_Param}.
\end{appendices}

\section*{Funding Declaration}
NGK appreciates financial support from the FAPESP--UNESCO-TWAS Project (Grant No. 2024/08346-8).
\section*{Acknowledgements}
HAC thanks FAPESP grant 2021/14335-0 of the ICTP--SAIFR for partial support.\\
\section*{Declarations}
\subsection*{Conflict of interest / Competing interests}
The authors declare that they have no conflict of interest and no competing interests.
\subsection*{Ethics approval and consent to participate}
Not applicable.
\subsection*{Data availability}
Data sets were not generated or analyzed during the current study. All results were obtained from numerical simulations using the original Python code developed by the authors.

\bibliographystyle{snstyle}
\bibliography{sn-bibliography}

\end{document}